\documentclass[journal=jctcce,manuscript=article]{achemso}

\usepackage[version=3]{mhchem} 
\mciteErrorOnUnknownfalse

\usepackage{amsmath,amssymb,amsthm, amsfonts, mathtools, dsfont} 
\usepackage{bbm,bm,tensor, braket}
\usepackage{eqnarray,array,enumerate}
\usepackage{csquotes}
\usepackage{siunitx}
\usepackage{booktabs}
\usepackage{graphicx,wrapfig, caption, float, subcaption, epstopdf, setspace}
\usepackage{hyperref}
\usepackage[section]{placeins}
\usepackage{multicol, multirow}
\usepackage{tikz,tkz-euclide}
\usetikzlibrary{shapes.geometric,arrows,arrows.meta, calc, positioning, automata, shadows, backgrounds,decorations.markings,decorations.pathreplacing}
\usepackage{tkz-fct}
\usetikzlibrary{intersections}
\usepackage{pgfplots}
\pgfplotsset{compat=1.9}
\usepgfplotslibrary{fillbetween}
\usepackage{pagecolor}\usepackage{amssymb}
\usepackage{algorithm, algpseudocode}
\usepackage{pdfpages}

\usepackage{cleveref}

\newcommand*{\citen}[1]{%
  \begingroup
    \romannumeral-`\x 
    \setcitestyle{numbers}%
    \cite{#1}%
  \endgroup   
}

\algrenewcommand\algorithmiccomment[1]{\hfill #1}

\def\br{\ensuremath\bm{r}}
\def\emp2{\ensuremath{E_{\text{corr}}^{\text{MP}2}}}

\def\ag{\ensuremath{\alpha}}
\def\bg{\ensuremath{\beta}}
\def\cg{\ensuremath{\gamma}}
\def\dg{\ensuremath{\delta}}

\author{Arno F{\"o}rster}
\email{a.t.l.foerster@vu.nl}
\affiliation{Theoretical Chemistry, Vrije Universiteit, De Boelelaan 1083, NL-1081 HV, Amsterdam, The Netherlands}
\author{Lucas Visscher}
\affiliation{Theoretical Chemistry, Vrije Universiteit, De Boelelaan 1083, NL-1081 HV, Amsterdam, The Netherlands}

\title{Low-order scaling $G_0W_0$ by pair atomic density fitting}


\begin{document}

\Large This article has been published in final form and open access by Journal of chemical theory and computation and can be accessed at
\href{https://pubs.acs.org/doi/10.1021/acs.jctc.0c00693}{https://pubs.acs.org/doi/10.1021/acs.jctc.0c00693}

\normalsize

\begin{abstract}
We derive a low-scaling $G_0W_0$ algorithm for molecules, using pair atomic density fitting (PADF) and an imaginary time representation of the Green's function and describe its implementation in the Slater type orbital (STO) based Amsterdam density functional (ADF) electronic structure code. We demonstrate the scalability of our algorithm on a series of water clusters with up to 432 atoms and 7776 basis functions and observe asymptotic quadratic scaling with realistic threshold qualities controlling distance effects and basis sets of triple-$\zeta$ (TZ) plus double polarization quality. Also owing to a very small prefactor, with these settings a $G_0W_0$ calculation for the largest of these clusters takes only 240 CPU hours. We assess the accuracy of our algorithm for HOMO and LUMO energies in the GW100 database. With errors of 0.24 eV for HOMO energies on the quadruple-$\zeta$ level, our implementation is less accurate than canonical all-electron implementations using the larger def2-QZVP GTO-tpye basis set. Apart from basis set errors, this is related to the well-known shortcomings of the GW space-time method using analytical continuation techniques as well as to numerical issues of the PADF-approach of accurately representing diffuse AO-products. We speculate, that these difficulties might be overcome by using optimized auxiliary fit sets with more diffuse functions of higher angular momenta. Despite these shortcomings, for subsets of medium and large molecules from the GW5000 database, the error of our approach using basis sets of TZ and augmented DZ quality is decreasing with system size. On the augmented DZ level we reproduce canonical, complete basis set limit extrapolated reference values with an accuracy of 80 meV on average for a set of 20 large organic molecules. We anticipate our algorithm, in its current form, to be very useful in the study of single-particle properties of large organic systems such as chromophores and acceptor molecules. 
\end{abstract}

\section{\label{sec:introduction}Introduction}\protect
Spectroscopy provides fundamental insights into the optical and electronic properties of matter and thus plays a decisive role in chemistry and material science.\cite{
Jensen2000,
Parson2007,
Martin-Drumel2016,
Sugiki2017,
Puzzarini2018} The great potential of computational spectroscopy is leveraged increasingly to complement and understand spectroscopic experiments.\cite{Choi2007,
Boukhvalov2008,
Zhang2009a,
Pedone2010,
Puzzarini2010,
Barone2011,
Berova2011,
Barone2012,
Kessler2018,
Puzzarini2019} Still, no existing computational method can be applied routinely to systems of hundreds of atoms and simultaneously predict the outcome of a spectroscopic experiment with satisfactory accuracy.\cite{Puzzarini2019} For ground state properties, Kohn-Sham (KS)\cite{Kohn1965} density functional theory (DFT)\cite{
Hohenberg1964,
Levy1979,
Engel2013} has been proven to be very accurate for many weakly correlated molecular systems.\cite{
Hickey2014,
Becke2014,
Pribram-Jones2015,
Yu2016a,
Goerigk2017a,
Mardirossian2017a,
Grimme2018} Excited particles, however, interact strongly with other electrons and semi-local or hybrid approximations to the exact functional of KS-DFT do not capture this physics correctly.\cite{Godby1988,
Engel1992,
Steinbeck2000,
Gruning2002,
Gruning2002a,
Malet2012,
Baerends2013,
VanMeer2014,
Gritsenko2016,
Baerends2017} Consequently, they fail to adequately describe single-particle excitations, being necessary to understand and predict phenomena like transport,\cite{Thygesen2007, Darancet2017, Thoss2018} tunneling,\cite{Unjic1991, Rignanese2001, Dial2012} or photoemission.\cite{Aryasetiawan1998,
Lischner2013,
Golze2018,
Golze2019,
Golze2020,
Kuhne2020a} 

Many body perturbation theory (MBPT)\cite{Luttinger1960,
Baym1961,Hedin1965} based on Hedin's equations describes the correlation of the excited electron with its surrounding by an expansion in powers of the response of the systems total classical potential to an external perturbation.\cite{Onida2002,
martin2016,
Reining2018} Only taking into account the first-order term of this expansion is called GW-approximation.\cite{Hedin1965,Hybertsen1986} Accounting for the major part of electron correlation,\cite{Deslippe2012, Reining2018, Golze2019, Pavlyukh2020} it makes MBPT computationally tractable, greatly improves over DFT for the description of single-particle excitations\cite{martin2016,Knight2016,Golze2019} and also paves the way toward accurate optical spectra using the Bethe-Salpeter equation formalism.\cite{Salpeter1951,Strinati1988} Large numbers of computational material science codes\cite{Deslippe2012,
Gonze2002,
Gonze2009,
Nguyen2012a,
Pham2013,
Kresse1996,
Kresse1996a,
Joubert1999,
FHIaims2009,
Blum2009,
Bruneval2016a,
Sangalli2019,
Kuhne2020a,
Schlipf2020} feature GW implementations and also in the quantum chemistry community it has acquired some momentum over the last years.\cite{
Blase2011,
Faber2011,
Strange2011,
Marom2012,
Baumeier2012,
Bruneval2013,
Caruso2013b,
Faber2013,
Korbel2014,
VanSetten2015,
Rangel2016,
Knight2016,
Li2016,
Duchemin2016,
Scherpelz2016,
Maggio2017,
Hung2017,
Olsen2019,
Bruneval2019,
Lewis2019,
Holzer2019,
Cazzaniga2020} 

The downside of the GW method is its huge operation count compared to KS-DFT, preventing its routine application to large systems. A popular approach to reduce the prefactor, frequently outperforming self-consistent approaches for charged excitations,\cite{Knight2016, Caruso2016} is the so-called $G_0W_0$-approximation in which the self-energy is calculated using a mean-field Green's function. Still, the operation count of a $G_0W_0$ calculation increases as $N^4$ as a function\bibnote{The notion of $N^{x}$ always implies $\mathcal{O}\left(N^x\right)$: That is, the function $f$, mapping the system size to the run-time of an algorithm, does asymptotically not grow faster than $N^{x}$.} of system size $N$ and compared to KS-DFT, GW quasi-particle (QP) energies converge significantly slower with the size of the single-particle basis.\cite{Shih2010, Nguyen2012, Golze2019} Consequently, the last years have witnessed some effort to reduce time-to-solution further which resulted in massively parallel implementations optimized for state-of-the-art supercomputers\cite{DelBen2019,Wilhelm2018} but also in notable algorithmic developments, including stochastic approaches,\cite{Neuhauser2014, Vlcek2017, Vlcek2018} implementations avoiding the explicit summation over empty electronic states in the polarizability, $P$\cite{Umari2010,Giustino2010, Lambert2013,Bruneval2016} low-rank approximations to the dielectric function $\epsilon$\cite{Nguyen2012, Pham2013, Govoni2015} or the screened interaction $W$,\cite{Rostgaard2010, Shao2016} and basis set error (BSE) correction schemes.\cite{Klimes2014, Riemelmoser2020, Loos2020a}

In \emph{ab initio} calculations on molecular systems, atom centered localized atomic orbitals (AO) are commonly employed\cite{Helgaker2014} In this representation, the dimensions of $W$ and $P$ grow as $N^2$, making the evaluation of $W$ an $N^6$ operation. One can employ an implicit low-rank approximation to both quantities by transforming them to a smaller auxiliary basis. Such transformations, most importantly density fitting (DF)\cite{
Billingsley1971,
Baerends1973,
Whitten1973,
Sambe1975,
Dunlap1979,
Dunlap1979a,
Vahtras1993,
Feyereisen1993,
Kendall1997,
Dunlap2000,
Dunlap2000a,
Weigend2006,
Dunlap2010a} and Cholesky decomposition (CD)\cite{Beebe1977,
Koch2003,
Aquilante2007,
Boman2008,
Aquilante2009} techniques, are employed in quantum chemistry since nearly half a century\cite{
Billingsley1971,
Baerends1973,
Whitten1973} and they are routinely used in GW implementations for molecules\cite{Foerster2011a,
Ke2011,
Ren2012,
VanSetten2013,
Caruso2013,
Kaplan2016,
Bruneval2016a,
Krause2017,
Tirimbo2020} where their accuracy is well documented.\cite{VanSetten2013, VanSetten2015} Using these techniques, the evaluation of $W$ becomes an $N^3$-operation with a sufficiently small prefactor. However, the transformations from product basis to auxiliary basis and back, usually implicit in the evaluation of $P$ and the self-energy $\Sigma$, respectively, still scale as $N^4$. 

This issue can in principle be avoided by constructing a sparse transformation matrix using local DF approximations (LDF).\cite{Watson2003,
Polly2004,
Sodt2006,
Sodt2008} However, conventional GW calculations are performed in frequency space, necessitating a representation of the Green's function in the MO basis where the sparsity of the transformation matrix is lost. From this perspective, the Green's function is more conveniently represented in imaginary time\cite{Almlof1991,
Haser1992,
Haser1993,
H.N.Rojas1995,
Rieger1999,
Steinbeck2000} since the energy denominator in $P$ factorizes and the relevant equations can be transformed to the AO basis where LDF might be used efficiently. 

LDF techniques have originally been proposed to evaluate the Fock matrix in generalized KS (gKS)- and Hartree-Fock (HF) calculations in a low-scaling fashion.\cite{Baerends1973} This is fortunate, since in imaginary time the evaluation of $\Sigma$ is equivalent to calculating the exact exchange contribution to the Fock matrix. In the most extreme LDF-variant, each AO-pair product is expanded in a set of auxiliary basis functions (ABF) centered on the same two atoms as the target pair of primitives. We refer to this approach as pair atomic DF (PADF) and note, that also the names concentric DF, pair-atomic resolution of the identity (PARI), and RI-LVL are encountered in the literature. It has been introduced by Baerends et al. in the 70s,\cite{Baerends1973} and subsequently employed in pure\cite{Guerra1998} and hybrid\cite{Watson2003, Krykunov2009} DFT calculations. As an efficient way to construct the Fock matrix, it has received renewed attention over the last years\cite{Merlot2013,
Hollman2014a,
Mejia-Rodriguez2014,
Manzer2015,
Lewis2016,
Hollman2017,
Lin2020} and its strengths and shortcomings for this task have been analyzed in detail.\cite{Rebolini2016,Wirz2017} It has also been applied to correlated methods and shown to be very accurate when appropriate auxiliary fit sets are used.\cite{Ihrig2015,Tew2018, Forster2020, Forster2020a} 

For the GW space-time approach\cite{H.N.Rojas1995} to be useful in practice, small grids not only in imaginary time but also in imaginary frequency as well as an efficient way to switch between both domains are needed to avoid potentially prohibitive prefactors and storage bottlenecks. How to address these technical issues has been shown by Kresse and coworkers\cite{Kaltak2014,Kaltak2014a} who subsequently presented cubic scaling GW implementations for periodic systems\cite{Liu2016, Grumet2018} and also low-scaling space-time RPA\cite{Wilhelm2016, Duchemin2019} and GW\cite{Wilhelm2018} implementations for molecular systems could be realized in the last years.
 
It has already been anticipated\cite{Ihrig2015} that PADF is especially well suited to implement GW in a low-scaling fashion. Against this background, we herein derive a GW space-time algorithm whose asymptotic cost associated with the calculation of $P$ and $\Sigma$ is reduced to $N^3$ independent of system size, and to $N^2$, when distance effects are exploited. We only discuss our $G_0W_0$-implementation here while self-consistent GW will be discussed in a future publication. However, we stress that within the herein presented framework quasi-particle\cite{VanSchilfgaarde2006, Kotani2007} and fully self-consistent GW is readily implemented as we always evaluate the complete self-energy matrix instead of only its diagonal in the MO basis.

We implemented our algorithm in the Slater type orbitals (STO) based Amsterdam density functional package (ADF).\cite{Snijders2001, ADF2019} Thus, our work is the first production level implementation of a GW method using STOs. While we do not aim at a comparison of different types of localized basis functions, we consider our implementation as a necessary first step towards a better understanding of the possible benefits of STOs in MBPT and also as a demonstration that they can be used efficiently in GW calculations. We also emphasize that the herein presented formalism is independent of the actual choice of basis functions, provided that they are local. We already note at this point that similar ideas have been presented by Wilhelm et al.\cite{Wilhelm2018} and implemented in the CP2K package.\cite{Kuhne2020a} We will start the following discussion in section~\ref{sec::theory} by defining the basic quantities in real space (RS) and imaginary time, discretize them using an AO basis and imaginary time grids and transform them to an auxiliary basis. From this starting point, we outline our algorithm and its implementation before we investigate its accuracy and computational efficiency in section~\ref{sec::Results}. Finally, section~\ref{sec::conclusion} concludes this work with a summary and perspectives on further research.

\section{\label{sec::theory}Theory}

\subsection{$G_0W_0$ in real space and imaginary time}
We start this section by briefly outlining the $G_0W_0$ approximation to Hedin's equations in the random phase approximation (RPA).\cite{martin2016} Using the molecular orbitals $\phi_n$ and corresponding orbital energies $\epsilon_n$ obtained from solving 
\begin{equation}
\label{genKS}
\left[
h^{(0)}(\br) - \epsilon_n
\right]\phi_n(\br) + \int_{\mathbb{R}^3} d \br' \ V_{xc}(\br,\br')\phi_n(\br') = 0 \;,
\end{equation} 
with a single-particle Hamiltonian $h^{(0)}$ and a potentially local exchange-correlation (xc) potential $V_{xc}$, the irreducible single-particle time-ordered Green's function in imaginary time is given as
\begin{equation}
\label{basicGreens}
G(\br,\br',i\tau) =  \Theta(\tau)\underline{G}(\br,\br',i\tau)
 -\Theta(-\tau)\overline{G}(\br,\br',i\tau) \;,
\end{equation}
with
 \begin{equation}
 \label{greensGW}
 \begin{aligned}
\underline{G}(\br,\br',i\tau) = & i\sum^{occ}_i \phi_i(\br)\phi^{\ast}_i(\br')e^{-|\epsilon_i - \epsilon_F| \tau} \;,\\
\overline{G}(\br,\br',i\tau) = & i\sum^{virt}_a \phi_a(\br)\phi^{\ast}_a(\br')e^{|\epsilon_a - \epsilon_F| \tau}
\end{aligned}
\end{equation}
being hole and particle Green's functions, respectively, $i,j, \dots$ ($a,b, \dots$) labeling occupied (virtual) orbitals, $\epsilon_F$ being the Fermi energy and $\Theta$ being the Heavyside step function. The independent-particle polarizability $P$ in the RPA is defined as 
\begin{equation}
\label{polarizability1}
P(\br,\br', i\tau) = 
-iG(\br,\br',i\tau)G(\br',\br,-i\tau) \;,
\end{equation}
and using \eqref{basicGreens} and \eqref{greensGW} can be written as
\begin{equation}
\label{polarizability}
P(\br,\br', i\tau) = 
-i \sum_{i}^{occ}\sum_a^{virt}
\phi_i(\br)\phi^{\ast}_i(\br')
\phi_a(\br')\phi^{\ast}_a(\br)
e^{-|\epsilon_a - \epsilon_F|\tau}
e^{-|\epsilon_i- \epsilon_F|\tau} \;.
\end{equation}
The polarizability is the kernel of a Dyson equation relating the reducible (or screened) Coulomb interaction $W(\br,\br',i\tau)$ to the bare Coulomb potential $V(\br,\br') = V'(\br,\br',i\tau)\delta(\tau-\nobreak\tau')$ (see e.g. ref. \citen{Stan2016} or \citen{Rostgaard2010}), 
\begin{equation}
W(\br,\br',i \tau - i\tau') = V(\br,\br') + \int d\tau_4 d\br_3 d\br_4 V(\br,\br_3)P(\br_3,\br_4, i\tau - i\tau_4) W(\br_4,\br',i\tau_4-i\tau') \;,
\end{equation}
which takes the simpler form\cite{Rieger1999}
\begin{equation}
\label{gwW}
W^{-1}(\br,\br',i \omega) =  V^{-1}(\br,\br') -  P(\br,\br', i\omega)
\end{equation}
in the imaginary frequency domain. 
From this quantity, the irreducible self-energy $\Sigma$ can be constructed\cite{martin2016} which is most conveniently split into a static and a dynamic contribution, $\Sigma = \Sigma^x + \Sigma^c$.
The former is the HF exchange kernel and is given as
\begin{equation}
\label{sigmaX}
\Sigma^x(\br,\br') 
=i\underline{G}(\br,\br',i\tau = 0)V(\br,\br') \;,
\end{equation}
and the latter as
\begin{equation}
\label{sigmaC}
\Sigma^c(\br,\br',i\tau) = iG(\br,\br',i\tau)\widetilde{W}(\br,\br',i\tau) \;,
\end{equation}
where we have introduced $\widetilde{W} = W - V$. 
In a self-consistent procedure, $G$ would be updated by solving another Dyson equation containing $\Sigma$ as its kernel. In a $G_0W_0$ calculation, $\Sigma^c$ is transformed to the imaginary frequency axis from where it is analytically continued to the complex plane.\cite{Cances2016,Han2017} The QP energy $\epsilon^{QS}_n$ is the $\omega$ which fulfills
\begin{equation}
\label{quasi-particle-equations}
0 = \omega - \epsilon_{n} - 
\braket{n| \operatorname{Re}\left(\Sigma^c(\omega)\right) + \Sigma_x - V_{xc}|n}
\;,
\end{equation}
where $\braket{n|O|m}$ denote matrix elements of an operator $O$ in the molecular orbital basis. 

\subsection{$G_0W_0$ in a local basis}
 
\paragraph{Discretization of real space}
Assuming we have represented imaginary time and frequency dependence of all quantities through suitable grids, we use (real) STOs $\chi$ to discretize RS, so that
\begin{equation}
\label{basisSetExpansion}
\phi_n(\br) = \sum_{\mu} b_{\mu n } \chi_{\mu}(\br) \;.    
\end{equation} 
Inserting this definition into \cref{basicGreens,greensGW} gives
\begin{align}
    \underline{G}(\br,\br',i\tau) = & \sum_i\sum_{\mu \nu} 
    \chi_{\mu}(\br)
    b_{\mu i } e^{-|\epsilon_i - \epsilon_F| \tau}b_{i \nu}
    \chi_{\nu}(\br') \\
    \overline{G}(\br,\br',i\tau) = & \sum_a\sum_{\mu \nu} 
    \chi_{\mu}(\br)
    b_{\mu a } e^{|\epsilon_a - \epsilon_F| \tau}b_{a \nu}
    \chi_{\nu}(\br') 
\end{align}
and from the identity
\begin{equation}
    G(\br,\br',i\tau) = \sum_{\mu \nu} 
    \chi_{\mu}(\br)
    G_{\mu \nu,\tau}
    \chi_{\nu}(\br')
\end{equation}
we obtain the representation of particle and hole Green's function in the STO basis,
\begin{align}
\label{greensGWbasisO}
    \underline{G}_{\mu \nu, \tau} = & \sum_i
    b_{\mu i } e^{-|\epsilon_i - \epsilon_F| \tau}b_{i \nu}\\
\label{greensGWbasisU}
    \overline{G}_{\mu \nu, \tau} = & \sum_a
    b_{\mu a } e^{|\epsilon_a - \epsilon_F| \tau}b_{a \nu} \;,
\end{align}
which for each discrete $i\tau$ can be seen as an energy-weighted density matrix.\cite{Surja2005} While $\Sigma$ also transforms as a 2-point correlation function,
\begin{equation}
\Sigma_{\mu\nu,\tau} = \int d \br d \br' 
\chi_{\mu}(\br)\Sigma(\br,\br', i\tau)\chi_{\nu}(\br') \;,
\end{equation}
all 2-electron operators transform as 4-point correlation functions,\cite{Starke2012}
\begin{align}
P_{\mu\kappa \nu \lambda, \tau} = & 
i\underline{G}_{\mu \nu, \tau}
\overline{G}_{\kappa \lambda,\tau} \\
V_{\mu \nu \kappa \lambda} = & 
\int d \br d \br' 
\chi_{\mu}(\br)\chi_{\nu}(\br)
V(\br,\br')
\chi_{\kappa}(\br')\chi_{\lambda}(\br') \\
\label{tildeW}
\widetilde{W}_{\mu \nu \kappa \lambda, \tau} = & 
\int d \br d \br' 
\chi_{\mu}(\br)\chi_{\nu}(\br)
\widetilde{W}(\br,\br',i\tau)
\chi_{\kappa}(\br')\chi_{\lambda}(\br') \;.
\end{align}
While in this representation $P$ is simply given as a Kronecker product, the calculation of the screened interaction \eqref{tildeW} from $P$ and $V$ requires the inversion of a matrix in the AO-product space $\mathcal{P} = \left\{\chi_{\mu}\right\} \otimes \left\{\chi_{\nu}\right\}$ for all frequency points (either of $W^{-1}$ as in \eqref{gwW} or of the dielectric function $\epsilon$ which is calculated from $P$ and $V$) whose dimension scales as $N^2$ with system size. Hence, the matrix inversion scales as $N^6$. 

This scaling does not reflect the systems physics and is simply an artefact of the chosen representation. The Eckard--Young theorem guarantees the optimal rank-$r$ approximation $M^{(r)}$ to some matrix $M$ to be given by the first $r$ terms in the sum on the \emph{r.h.s.} of
\begin{equation}
\label{eckardYoung}
M^{(r)} = \sum_i^{r} \sigma_i v_i \otimes u_i, \quad \sigma_i \geq \sigma_{i + 1} \;,
\end{equation}
where $\sigma$ is a singular value and $v_i$ and $u_i$ are vectors of the matrices $V$ and $U$ from the singular value decomposition (SVD) of $M$. In this way one can indeed show that the ranks of $P$, $V$ and $W$ should only grow linearly with system size\cite{Schutski2017} and using \eqref{eckardYoung} one might decompose $P$, $V$ and $W$ (given that they are symmetric) as
\begin{equation}
\label{decompose}
M_{\mu \nu\kappa\lambda} = \sum_{pq} C_{ \mu \nu p}Z_{pq}  [C^{T}]_{q \kappa \lambda }, \quad M = P,V,\widetilde{W} \;,
\end{equation}
where $Z$ is the diagonal matrix of singular values and C collects the left singular vectors of $M$. An explicit SVD would scale as $N_{AO}^4 r$ and is prohibitive in practice.\cite{Schutski2017} Instead, it is common practice to represent $V$ and $W$ in a predefined auxiliary basis $\mathcal{A} = \left\{f\right\}$, growing linearly with system size. Expanding all AO-pair products in terms of $\mathcal{A}$,
\begin{equation}
\label{pair_density_expansion}
\chi_{\mu}(\br)\chi_{\nu}(\br) = \sum_{p} C_{\mu \nu p} f_{p}(\br) \;,
\end{equation}
where Greek lowercase letters label AOs and the Roman lowercase letters $p,q, \dots$ refer to ABFs, $V$ and $\widetilde{W}$ can be expressed as
\begin{align}
V_{p q} = & 
\int d \br d \br' 
f_{p}(\br) V(\br,\br')f_{q}(\br') \\
\widetilde{W}_{p q} = & 
\int d \br d \br' 
f_{p}(\br) \widetilde{W}(\br,\br')f_{q}(\br') \;,
\end{align}
and with \eqref{decompose} and \eqref{pair_density_expansion}, the equations to be solved in a $G_0W_0$ calculation become
\begin{align}
\label{pFitting}
P_{pq,\tau} = & C_{\mu \nu p}P_{\mu \nu\kappa \lambda, \tau} C_{\kappa \lambda q} = 
-i C_{\mu \nu p}\underline{G}_{\mu \kappa, \tau}
\overline{G}_{\nu \lambda, \tau} C_{\kappa \lambda q} \\
\label{wFitting}
W_{pq,\omega} = & V_{pq} + V_{pr} P_{rs,\omega} W_{sq,\omega} 
= \left[V^{-1} - P\right]_{pq, \omega}^{-1} \\
\label{sxFitting}
\Sigma^x_{\mu \nu} =&
i\sum_{\kappa \lambda} \sum_{pq}
\underline{G}_{\kappa \lambda,\tau=0}
C_{\mu \kappa q}
V_{pq}
C_{\nu \lambda p } \\
\label{scFitting}
\Sigma^c_{\mu \nu, \tau} =&
i\sum_{\kappa \lambda} \sum_{pq}
G_{\kappa \lambda, \tau}
C_{\mu \kappa q}
\widetilde{W}_{pq, \tau}
C_{\nu \lambda q} \;,
\end{align}
replacing \cref{polarizability,gwW,sigmaX,sigmaC}. In this set of equations, \eqref{pFitting} is the computational bottleneck. While the basis transformation in the first equation in \eqref{pFitting} would scale as $N^5$, also using the second equation one ends up with a scaling of $N^4$. The same is actually true for \eqref{sigmaX} and \eqref{sigmaC}, however, as in a $G_0W_0$ calculation only the diagonal elements of $\Sigma$ in the MO basis are needed, the computational effort reduces to $N^3$. 

Improvements over the $N^4$-scaling can be achieved in essentially two ways. The first way relies on the asymptotically exponential decay of the density matrix.\cite{Kohn1996, Baer1997, Goedecker1999} Ochsenfeld and coworkers exploited the resulting sparsity in $\underline{G}$ and $\overline{G}$\cite{Schindlmayr2000} to calculate correlation energies in second order M{\o}ller--Plesset perturbation theory (MP2)\cite{
Zienau2009,
Maurer2014,
Maurer2014a} and RPA.\cite{Schurkus2016,
Luenser2017,
Graf2018,
Graf2019} It is an obvious drawback of the approach that in 3D systems the density matrix is less sparse as one would hope for,\cite{Rudberg2008,Rudberg2011,Vandevondele2012} especially for large AO basis sets with many diffuse functions commonly employed in GW calculations. The second way is to construct a sparse map from $\mathcal{P}$ to $\mathcal{A}$. How this can be achieved will be discussed in the next paragraph.

\paragraph{\label{par::PADF}Local density fitting approximations}

Given some target precision $\epsilon$, the two main goals of DF are first, to find a matrix $M'$ with dimension $N_{aux}$ for which 
\begin{equation}
\label{targetprecision}
\|M-M'\| < \epsilon
\end{equation}
with $N_{aux}$ as small as possible and $M$ defined by \eqref{decompose} and second, to improve over the unfavourable scaling of \cref{pFitting,sxFitting,scFitting} by constructing $C$ in a way that it becomes sparse. Both goals are in principle in conflict with each other. In DF, one minimizes the residual function
\begin{equation}
\label{resdual}
r_{\mu \nu}(\br) = \chi_{\mu}(\br)\chi_{\nu}(\br) -\sum_p C_{\mu \nu p} f_p(\br) \quad \forall \mu,\nu \;,
\end{equation}
with respect to some appropriate norm. In the RI-V approach, the Coulomb repulsion of $r$ is minimized,
\begin{equation}
    \frac{\partial}{\partial C_{\kappa \lambda q}}\int d \br d\br'\; r_{\kappa \lambda}(\br) V(\br,\br') r_{\mu \nu}(\br') = 0 \;,
\end{equation}
and it follows that
\begin{equation}
\label{riV}
   \sum_{p} C_{\mu \nu p}V_{pq} = \int d \br d\br'\; 
    \chi_{\mu}(\br)\chi_{\nu}(\br) V(\br,\br') f_q(\br') \;,
\end{equation}
i.e. the error in the low-rank approximation of $V$ is quadratic in $r$ since the terms linear in $C$ vanish. Of course, a similar conclusion can not be drawn for $P$ and consequently also not for $W$. Still, it seems that this metric is an excellent choice if the goal is to fulfil \eqref{targetprecision} with $N_{aux}$ as small as possible and using auxiliary fit sets from standard libraries. As shown by van Setten et al, QP HOMOs and LUMOs only deviate by a few meV from the ones obtained from calculations without any low-rank approximation\cite{VanSetten2013, VanSetten2015} when appropriate auxiliary fit sets\cite{Eichkorn1995, Weigend2006} are used.

On the other hand, RI-V is a very bad choice in the sense that the slow decay of the kernel of the Coulomb operator ensures that $C$ will be dense. In the RI-SVS approach,\cite{Dunlap1979, Feyereisen1993} \eqref{resdual} is minimized with respect to the $L_2$ norm which requires larger $N_{aux}$ to fulfil \eqref{targetprecision} but results in a $C$ with the number of non-zero elements increasing only linearly with system size for exponentially decaying basis functions. It has been shown by Wilhelm et al. that this approach results in tremendous speed-ups in the evaluation of \cref{pFitting,sxFitting,scFitting} without requiring to large $N_{aux}$ to make the evaluation of \eqref{wFitting} problematic for systems of more than 1000 atoms.\cite{Wilhelm2018} However, for rather small systems with a 3D structure, the number of non-zero elements in $C$ will not be much different from $N_{AO}^2 \times N_{aux}$. Thus, due to the usually larger $N_{aux}$ compared to RI-V, the method will only be advantageous for sufficiently large systems.\cite{Wilhelm2018} 

In LDF approximations, this shortcoming is addressed by building in sparsity into the fitting procedure \emph{a priori}. In PADF, an expansion of the pair-density $\chi_{\mu}(\br)\chi_{\nu}(\br)$ of the form
\begin{equation}
\chi_{\mu}(\br)\chi_{\nu}(\br) = \sum_{p \in A \cup B }C_{\mu \nu p} f_p(\br) \quad \forall \mu \in A, \nu \in B \;
\end{equation}
is employed so that the number of non-zero elements in $C$ scales at most quadratic with system size. In our implementation, we also define thresholds $d_{\mu\nu}$ for each AO-product and assume $C_{\mu \nu p} = 0$ if $|\bm{R}_A - \bm{R}_B| > d_{\mu \nu}$ so that the number of non-zero elements in $C$ only increases linearly.\cite{Forster2020} For each atom, we also reorder all AOs from the most diffuse to the least diffuse one so that all non-zero elements in $C$ are grouped in dense blocks. Eq. \eqref{resdual} becomes
\begin{equation}
\label{res_PADF}
r^{PADF}_{\mu \nu}(\br) = \chi_{\mu}(\br)\chi_{\nu}(\br) - \sum_{p \in A \cup B } C_{\mu \nu p} f_p(\br) \quad \forall \mu \in A, \nu \in B \;,
\end{equation}
which is minimized with respect to the Coulomb metric. Solving 
\begin{equation}
    \frac{\partial}{\partial C_{\kappa \lambda q}}\int d \br d\br'\; r^{PADF}_{\kappa \lambda}(\br) V(\br,\br') r^{PADF}_{\mu \nu}(\br') = 0 \;,
\end{equation}
does not lead to an equation of the form \eqref{riV} as the terms linear in $C$ do not vanish. Thus, determining $C$ by solving \eqref{riV} for all (nearby) atom pairs $(A,B)$ leads to errors for $V$ linear in $r$ (the same holds for DF in the RI-SVS approach). It has been concluded that the resulting errors are too large for the method to be useful in HF calculations.\cite{Hollman2017,Dunlap2000,Dunlap2000a} This might be true when standard auxiliary fit sets are used which are optimized for global DF. In principle, the error of the expansion \eqref{pair_density_expansion} can always be made arbitrary small when an appropriate fit set is used although this is highly non trivial. Simply increasing the number of ABFs does not always result in reduced errors and might even lead to numerical instabilities in the fitting procedure due to an increase of linear dependencies in the auxiliary basis.\cite{Forster2020} 

Another difficulty arises from the presence of diffuse functions in the AO-basis set. To understand the source of the problem, we recall that very large AO basis sets with many diffuse functions might be locally overcomplete which causes almost linear dependence of a subset of basis functions. These lead to numerical instabilities in the SCF\cite{Klahn1977} during canonical orthonormalization when the condition number of the AO-overlap matrix approaches infinity.\cite{Lowdin1967} To restore numerical stability, one projects out the almost linearly dependent part from the basis by removing eigenvectors from the transformation matrix corresponding to eigenvalues of the AO-overlap matrix smaller than some threshold $\epsilon_D$,\cite{Kudin2000} effectively diminishing the basis set size. This is not a severe restriction in practice since numerical instabilities usually do not occur when all eigenvalues are larger than $\epsilon_D = 10^{-6}$ - $10^{-7}$.\cite{Suhai1982, Kudin2002, Lehtola2020}  

Using PADF, numerical instabilities can already occur when all eigenvalues are considerably larger as has e.g. been observed for linear-response TDDFT with augmented basis sets\cite{Schipper2000} and MP2/QZ calculations\cite{Forster2020}. The reason for this behaviour is that individual fitting coefficients can become quite large for diffuse products from AOs centered on distant atoms. Note, that this is a fundamental difference to global DF. As a qualitative example, consider a linear alkane chain $\text{C}_n\text{H}_{2n+2}$ and the pair product of a diffuse AOs on \ce{C1} and $\text{C}_n$, respectively. The AOs will only have some (small) overlap in the middle of the chain. In global DF, this pair product could possibly be described very well with only a small set of ABFs centered on atoms in this region. In PADF, this overlap needs to be described with the asymptotic tails of diffuse ABFs on \ce{C1} and $\text{C}_n$. When there is no appropriate ABF in the auxiliary basis, this will lead to very large fitting coefficients for some (diffuse) ABFs. In the transformation of the Coulomb potential from auxiliary basis to AO-product basis, these large fitting coefficients must cancel with contributions with opposite sign which is numerically unstable.\cite{DeJong2002} Thus, relatively small errors might accumulate during the SCF and lead to an erroneous (hole) density matrix and potentially wrong eigenvalues.

To summarise, projecting out parts of the basis during canonical orthonormalization plays a dual role when PADF is used in the SCF. First, it ensures numerical stability of the SCF and second, as a side-effect, it removes the part of the basis which potentially results in diffuse AO-products which are potentially difficult to fit. This nicely illustrates that the appropriate choice of auxiliary basis and the problem of linear dependencies are intertwined. Adding more diffuse functions to the auxiliary basis the pair product in our example can be better approximated, the fitting coefficient become smaller, and the linear dependency problem is extenuated. This means, the number of AOs which needs to be removed becomes smaller and larger basis sets can be used in practice. 

In the present work, we employ auxiliary fit sets which have been optimized for gKS calculations with PADF. Using these fit sets, we have shown recently\cite{Forster2020,Forster2020a} that the accuracy of PADF-MP2 is similar to global DF-MP2 with GTOs for basis sets of up to TZ quality. On the other hand, using quadruple-$\zeta$ (QZ) and also smaller basis sets augmented with diffuse functions results sometimes in unreliable PADF-MP2 ground state energies. It is clear that the same issues will arise in $GW$ calculations.

For correlated methods we observed, that a value of $\epsilon_D = 10^{-3}$, corresponding to a drastic truncation of the basis, seems to provide a good trade-off between accuracy and numerical stability for all basis sets beyond TZ quality and also augmented basis sets. However, while this truncation prevents collapse to artificially low QP energies, it also leads to deteriorated results compared to the default of $\epsilon = 10^{-4}$. Increasing the basis set more and more, larger and larger parts of the virtual space need to be projected out which ultimately prevents us from reaching the complete basis set (CBS) limit for correlated methods. We expect, however, that carefully optimized auxiliary fit sets will enable the numerically stable application of PADF to these methods with larger basis sets. Before we discuss the accuracy of the present approach in section~\ref{sec::Results}, we will describe in some detail how PADF can be used to implement \cref{pFitting,wFitting,sxFitting,scFitting} efficiently.

\subsection{GW equations with pair atomic density fitting}

In this section we outline how the sparsity of the map from $\mathcal{P}$ to $\mathcal{A}$ can be exploited to implement GW in a low-scaling fashion.  

\paragraph{Imaginary time and frequency grids}
After calculation of the Coulomb potential and its inverse in the basis of ABFs and the basis transformation matrix $C$ as described in section~\ref{par::PADF}, we calculate imaginary  frequency and imaginary time grids,
$\left\{\omega_k\right\}_{k = 1, \dots N_{\omega}}$, $\left\{\tau_k\right\}_{k = 1, \dots N_{\tau}}$, respectively. As outlined by Kresse an coworkers,\cite{Kaltak2014} they can be evaluated by minimizing either the $L_{\infty}$ (Chebyshev) or $L_2$ norm of 
 \begin{equation}
 \label{errorDistributionFunction}
 \eta\left(\left\{\alpha,\beta\right\},x\right) = \frac{1}{x} - f\left(\left\{\alpha,\beta\right\},x\right) \;, \quad f = 
 \begin{cases}
 \displaystyle
 2\sum_{k = 1}^{N_{\tau}}\alpha_k e^{-\beta_k x} & \beta = \tau
 \\
 \displaystyle
 \frac{1}{\pi} \sum_{k=1}^{N_{\omega}} \alpha_k\frac{2x}{x^2 + \beta_k^2}
 & \beta = \omega
 \end{cases}
 \end{equation}
 with respect to the parameter sets $\alpha$,$\beta$, where $x \in [\epsilon_{min},\epsilon_{max}]$, where $\epsilon_{min}$ ($\epsilon_{max}$) denotes the smallest (largest) KS orbital energy difference. Imaginary time and imaginary frequency domain are connected through Laplace transforms (See also Cancés et al.\cite{Cances2016}),
\begin{align}
\label{fft1}
f(i\tau) = & \frac{i}{2 \pi} \int d\omega f(i \omega) (\cos({\omega\tau}) + i\sin ({\omega\tau}) )\\
\label{fft2}
f(i\omega) = & -i \int d\tau f(i \tau) (\cos({\omega\tau}) - i\sin ({\omega\tau}) ) \;.
\end{align}
For our purpose, it is sufficient to treat them as Fourier transforms. To avoid potentially inaccurate interpolation to equidistant grids in order to use discrete Fourier transforms, we discretize \eqref{fft2} as
 \begin{equation}
 \label{exactFFT}
 f(i\omega_k) =
 -i\sum^{N_{\tau}}_{j}\left\{\gamma^{(c)}_{kj}\cos (\omega_k \tau_j) 
 \left(f(i \tau_j) + f(-i \tau_j)\right) - 
 i \gamma^{(s)}_{kj}\sin (\omega_k \tau_j) 
 \left(f(i \tau_j) - f(-i \tau_j)\right)\right\} \;,
 \end{equation}
where the weights $\gamma^{(c)}_{kj}$ and $\gamma^{(s)}_{kj}$ account for the non-uniformity of the grids. They are chosen to minimize the $L_2$ norm of the error introduced by \eqref{exactFFT} for $f (i\tau) = e^{-x |\tau|}, x \in [\epsilon_{min},\epsilon_{max}]$, with respect to the exact transformation eq. \eqref{fft2}. By inverting the matrices $\gamma^{(c)}_{kj}\cos (\omega_k \tau_j) $ and $\gamma^{(s)}_{kj}\sin (\omega_k \tau_j) $, respectively, one can use the same relation to transform $f$ from imaginary frequency to imaginary time. To calculate the imaginary time grid, we minimize the $L_{\infty}$ norm of \eqref{errorDistributionFunction} as implemented by Helmich-Paris et al.\cite{Takatsuka2008a, Helmich-Paris2016} and in imaginary frequency we minimize the $L_2$ norm on a logarithmic grid using a Levenberg-Marquardt algorithm.\cite{Levenberg1944, Marquardt1963} Both algorithms require pretabulated values to converge to an acceptable local minimum. For the imaginary time domain, we use the values distributed with the source-code of Helmich-Paris et al.\bibnote{
 Source code available on \href{https://github.com/bhelmichparis/laplace-minimax}{https://github.com/bhelmichparis/laplace-minimax}} and for the imaginary frequency domain we tabulated our own values which we include in the supporting information.

\paragraph{Polarizability}
After the Green's function \eqref{greensGWbasisO} and \eqref{greensGWbasisU} have been constructed, $P$ can be evaluated. In this section, we use $\mu,\nu,\kappa,\lambda$ to denote AOs, $\alpha, \beta, \gamma, \delta$ to denote ABFs, and the convention that $(\mu, \alpha) \in A$, $(\nu, \beta) \in B$, $(\kappa ,\gamma) \in C$, $(\lambda, \delta) \in D$, where $A,B,C,D$ label atoms. We denote the three-leg tensor collecting all fitting coefficients corresponding to all products formed from AOs centred on $A$ and $B$ and to ABFs centred on $B$ as $C^{ABB}$, i.e. $C^{ABB}$ contains only coefficients corresponding to fit-functions centred on $B$. Consequently, the fitting coefficient tensor corresponding to all products formed from AOs on $A$ and $B$ and to ABFs centred on $A$ and $B$ is split into $C^{ABB} + C^{BAA}$. We also define $C^{ABB}= \frac{1}{1+\delta_{AB}}\widetilde{C}^{ABB}$ to avoid complications from double-counting. The contribution of each atom pair $(A,B)$ to $P$, eq. \eqref{pFitting}, is given as the sum of four contributions
\begin{equation}
\label{pdecompose}
P^{AB}_{\alpha \beta, \tau} = 
-i \left(
P^{AB,I}_{\alpha \beta, \tau} +
P^{AB,II}_{\alpha \beta, \tau} + 
P^{AB,III}_{\alpha \beta, \tau} + 
P^{AB,IV}_{\alpha \beta, \tau}  \right) \;,
\end{equation}
where
\begin{equation}
\begin{aligned}
\label{pIntermediates}
P^{AB,I}_{\alpha \beta, \tau} = &
\sum_{\mu\nu\kappa \lambda}
C^{DAA}_{ \lambda \mu \alpha} 
\underline{G}^{DC}_{\lambda \kappa, \tau}
\overline{G}^{AB}_{\mu \nu, \tau} 
C^{CBB}_{\kappa \nu \beta} \\
P^{AB,II}_{\alpha \beta, \tau} = &
\sum_{\mu\nu\kappa \lambda}
C^{DAA}_{ \lambda \mu \alpha} 
\underline{G}^{AC}_{\mu \kappa, \tau}
\overline{G}^{DB}_{\lambda \nu, \tau} 
C^{CBB}_{\kappa \nu \beta} \\
P^{AB,IV}_{\alpha \beta, \tau} = &
\sum_{\mu\nu\kappa \lambda}
C^{DAA}_{ \lambda \mu \alpha} 
\underline{G}^{AB}_{\mu \nu, \tau}
\overline{G}^{DC}_{\lambda \kappa, \tau} 
C^{CBB}_{\kappa \nu \beta} \;.
\end{aligned}
\end{equation}
As $\underline{G}$ and $\overline{G}$ are symmetric, the symmetry of the Kronecker product implies that $P$ is symmetric as well and consequently $P^{AB,III} = \left[P^{BA,II}\right]^T $ and $P^{AB} = \left[P^{BA}\right]^T $. Also note, that $\operatorname{Re}\left( P\right) = 0$. Defining the intermediates
\begin{align}
\label{Fupper}
\underline{F}^{ABB}_{\mu \nu \beta, \tau} = & 
\sum_{\kappa} \underline{G}^{AC}_{\mu \kappa, \tau}
C^{CBB}_{ \kappa \nu \beta} \\
\label{Flower}
\overline{F}^{ABB}_{\mu \nu \beta, \tau} = & 
\sum_{\kappa} 
\overline{G}^{AC}_{\mu \kappa, \tau} 
C^{CBB}_{\nu' \nu \beta}  \\
\label{Hlower}
\underline{H}^{ACB}_{\mu \kappa \beta, \tau} = &
\sum_{\nu}
\underline{F}^{ABB}_{\mu \nu \beta, \tau}
\overline{G}^{BC}_{\nu \kappa, \tau}  \\
\label{Hupper}
\overline{H}^{ACB}_{\mu \kappa \beta, \tau} = & 
\sum_{\nu}
\overline{F}^{ABB}_{\mu \nu \beta, \tau}
\underline{G}^{BC}_{\nu \kappa, \tau} \;,
\end{align}
\eqref{pIntermediates} is most conveniently evaluated as 
\begin{equation}
\begin{aligned}
\label{pdecompose2}
P^{AB,I}_{\alpha \beta, \tau}   + P^{AB,IV}_{\alpha \beta, \tau} 
= &
\sum_{\nu\kappa}\left(
\underline{H}^{CBA}_{\kappa \nu \alpha, \tau} +
\overline{H}^{CBA}_{\kappa \nu \alpha, \tau}
\right) C^{CBB}_{\kappa \nu \beta} \\
P^{AB,II}_{\alpha \beta, \tau} = &
\sum_{\mu\nu}
\overline{F}^{BAA}_{\nu \mu \alpha, \tau} 
\underline{F}^{ABB}_{\mu \nu \beta, \tau} \;.
\end{aligned}
\end{equation}

We parallelize the outermost loop over all atoms and perform all tensor contractions using level-3 BLAS. No step involves more than three atomic centers and since tensor contractions corresponding to distant centers (for which all elements in C are zero) can be skipped, the operation count scales asymptotically as $N^2$. We always evaluate the intermediates \cref{Fupper,Flower,Hupper,Hlower} on the fly since storage of 2-center quantities with more than 2 indices would quickly become prohibitive. 
\paragraph{Screened Coulomb interaction}
After having evaluated $P$ for all atom pairs, $\widetilde{W}$ can be evaluated as in conventional approaches using matrices of dimension $N_{aux} \times N_{aux}$. After transforming the matrix $P$ (which is even in imaginary time) to the imaginary frequency axis using \eqref{exactFFT}, the screened interaction $\widetilde{W}_{\omega}$ is obtained by inversion,
\begin{equation}
\label{Wfrequency}
\widetilde{W}_{\omega} = \left[V^{-1} - P_{\omega}\right]^{-1} - V\;.
\end{equation}
For all $\omega$, $W$ is stored in distributed memory. Note, that on the imaginary frequency axis, $\operatorname{Im}\left(P\right) = 0$ and thus $\operatorname{Im}\left(\widetilde{W}\right) = 0$ as well. To evaluate \eqref{Wfrequency}, the dielectric function is not constructed explicitly as it would not be symmetric and its inversion would be computationally demanding. We invert $V^{-1} - P_{\omega}$ (and  $V$ which only needs to be done once) using an LU decomposition with partial pivoting as implemented in SCALAPACK. Note, that inversion using CD would be numerically unstable since $C$ might not be full-rank and thus does not necessarily conserve positive semi-definiteness. We subsequently transform $\widetilde{W}$ back to imaginary time.

\paragraph{Self-energy}
Next, the contributions to $\Sigma$ for all atom pairs,
\begin{equation}
\label{sigmaAll}
\Sigma^{c,AB}_{\mu \nu, \tau} = 
i \left(\Sigma^{AB,I}_{\mu \nu, \tau} +
\Sigma^{c,AB,II}_{\mu \nu, \tau} + 
\Sigma^{c,AB,III}_{\mu \nu, \tau} + 
\Sigma^{c,AB,IV}_{\mu \nu, \tau}  \right) \;,
\end{equation}
are evaluated, where $\Sigma^{AB,III} = \left[ \Sigma^{BA,II}\right]^T $, $\Sigma^{AB} = \left[ \Sigma^{BA}\right]^T $. Also, $\operatorname{Re}\left( \Sigma^c\right) =0$, since $\operatorname{Re}\left( G \right) =0$ and $\operatorname{Re}\left(\widetilde{W}\right) =0$. We only give here the equations for $\Sigma^c(i\tau)$ as $\Sigma^x$ is obtained in exactly the same way by replacing $\widetilde{W}$ with $V$ and using $\underline{G}(i\tau = 0)$. As $\Sigma$ is an uneven function in imaginary time, we also need to evaluate $\Sigma(-i\tau)$ to be able to Fourier transform it to the imaginary frequency axis. The corresponding equations can be retrieved from the ones for $\Sigma(i\tau)$ by simply exchanging $\underline{G}$ with $\overline{G}$ and replacing upper bars with lower bars in all intermediates. To express the individual contributions to $\Sigma$ we define the intermediate
\begin{equation}
\label{intermediateI}
I^{ABC}_{\mu \nu \gamma,\tau} = C^{ABB}_{\mu \nu \beta} \widetilde{W}^{BC}_{\beta\gamma,\tau} \;,
\end{equation}
and together with \eqref{Flower} and \eqref{Fupper} we obtain
\begin{align}
\label{sigmaI}
\underline{\Sigma}^{c,AC,I}_{\mu \kappa, \tau} = &
\sum_{\nu \lambda} \sum_{\alpha \gamma}
\underline{G}^{DB}_{\lambda \nu, \tau}
C^{DAA}_{\lambda \mu  \ag}
\widetilde{W}^{AC}_{\ag \cg, \tau}  
C^{BCC}_{\nu \kappa  \cg}
= \sum_{\nu \ag}  
\underline{F}^{BAA}_{\nu \mu \ag, \tau} 
I^{BCA}_{ \nu \kappa \ag,\tau} \\
\label{sigmaII}
\underline{\Sigma}^{c,AC,II}_{\mu \kappa, \tau} = &
\sum_{\nu \lambda} \sum_{\alpha \beta}
\underline{G}^{DB}_{\lambda \nu, \tau}
C^{DAA}_{\lambda \mu  \ag}
\widetilde{W}^{AB}_{\ag \bg, \tau}  
C^{CBB}_{\kappa  \nu \bg}
= \sum_{\nu \ag}  
\underline{F}^{BAA}_{\nu \mu \ag, \tau}
 I^{CBA}_{\kappa \nu \ag,\tau} \\
 \label{sigmaIII}
\underline{\Sigma}^{c,AC,IV}_{\mu \kappa, \tau} = &
\sum_{\nu \lambda} \sum_{\delta \beta}
\underline{G}^{DB}_{\lambda \nu, \tau}
C^{ADD}_{\mu \lambda  \dg}
\widetilde{W}^{DB}_{\dg \bg, \tau}  
C^{CBB}_{\kappa  \nu \bg}
= \sum_{\lambda\delta}  
\left[
\sum_{\nu }
\underline{G}^{DB}_{\lambda \nu, \tau}
I^{CBD}_{\kappa \nu \dg,\tau}
\right]
C^{ADD}_{\mu \lambda  \dg} \;.
\end{align}
As for $P$ we parallelize the outermost loop over all atoms and completely rely on level-3 BLAS for all tensor contractions. Due to its prefactor of $N^2_{AO, l} \times N^2_{aux, l}$, where $N_{aux,l}$ ($N_{AO, l}$) denote the number of ABFs (AOs) on on atomic center, the calculation of $I$ is the most expensive step. The asymptotic operation count can be reduced significantly as the screened interaction $\widetilde{W}$, unlike the Coulomb interaction, decays exponentially as direct consequence of the exponential decay of the Green's function. In our current implementation, we do not fully exploit this property as we essentially treat $\widetilde{W}$ like the bare Coulomb potential in the calculation of $\Sigma^x$. In the same way as for $C$, we can skip all tensor contractions for approximately non-Coulomb-interacting atom pairs. For weakly interacting pairs, we rely on multipole expansions of the Coulomb potential to reduce the prefactor of all contractions involving $W$ (and $V$ for $\Sigma^x$) considerably so that $\Sigma$ can also be evaluated with quadratic operation count. Fully exploiting the exponential decay of $\widetilde{W}$, the asymptotic scaling can possibly be reduced further.

\paragraph{Quasi-particle equations}
$\Sigma^c$ is subsequently transformed to the MO basis and its diagonal elements to imaginary frequency space. With \eqref{exactFFT},
\begin{equation}
\label{fft3}
\Sigma^c_{nn,\omega_k} = 
 -i\sum_{j}\gamma^{(c)}_{kj}\cos (\omega_k \tau_j)
 \left[
 \underline{\Sigma}^c_{nn,\tau_j} +
 \overline{\Sigma}^c_{nn,\tau_j} 
 \right]
 - \sum_{j} \gamma^{(s)}_{kj}\sin (\omega_k \tau_j)
 \left[
 \underline{\Sigma}^c_{nn,\tau_j} -
 \overline{\Sigma}^c_{nn,\tau_j} 
 \right]\;,
\end{equation}
from which the QP equation \eqref{quasi-particle-equations} is solved. We analytically continue (AC) $\Sigma^c_{nn}$  to the real frequency axis using a Padé-approximant of order $N_{\omega}$ as described by Vidberg and Serene\cite{Vidberg1977} and solve \eqref{quasi-particle-equations} for all states of interest using bisection. While the present approach it not applicable to core level excitations,\cite{Golze2018, Golze2019, Golze2020} it predicts QP energies in the valence region with good accuracy in case the QP solution is sufficiently distant from any pole of the self-energy.\cite{Ke2011, Wilhelm2016a, Govoni2018, Golze2019} This is always the case for molecules with a large KS HOMO-LUMO gap. Note, that in these cases small imaginary frequency grids are sufficient to ensure good accuracy for particle and hole states in the valence region. 

To summarize this section, a pseudocode of our implementation together with theoretical asymptotic scaling with system size is given in figure~\ref{alg::GW}.

\begin{algorithm}[!hbt] 
\begin{algorithmic} 
\State Input MO coefficients $b_{\mu n}$, orbital energies, $\epsilon_n$ from \eqref{genKS}
\State Compute $C$, $V$, $V^{-1}$
\State Compute $\left\{\tau_i\right\}_{i = 1, \dots, N_{\tau}}$, $\left\{\omega_k\right\}_{k = 1, \dots, N_{\omega}}$, $\left\{\gamma^{(c)}_{ki}, \gamma^{(s)}_{ki}\right\}_{k = 1, \dots, N_{\omega}, i = 1, \dots N_{\tau}}$
\State
 \For {$\tau = \tau_1, \tau_2, \tau_3 \dots ,  \tau_{N_{\tau}}$}
   \State Calculate $G$ using \eqref{greensGWbasisO}, \eqref{greensGWbasisU}
   \Comment{$N^3 N_{\tau}\phantom{N_{\omega}}$}
   \For {$A \in N_{atom}, B \in N_{atom}$}
      \State Evaluate $P^{AB}(\tau_i)$ using \eqref{pdecompose}-\eqref{pdecompose2} \Comment{$N^2 N_{\tau}\phantom{N_{\omega}}$}
 \EndFor
      \For {$\omega = \omega_1, \omega_2, \omega_3 \dots ,  \omega_{N_{\omega}}$} 
      \State Calculate contribution to $P(\omega_i)$ using \eqref{exactFFT} \Comment{$N^2N_{\tau} N_{\omega}$}
  \EndFor
\EndFor
\For {$\omega = \omega_1, \omega_2, \omega_3 \dots ,  \omega_{N_{\omega}}$}
  \State Calculate $W(\omega_k)$ using \eqref{Wfrequency}
  \Comment{$N^3 N_{\omega}\phantom{N_{\tau}}$}
\EndFor
\For {$\tau = \tau_1, \tau_2, \tau_3 \dots ,  \tau_{N_{\tau}}$}
\For {$\omega = \omega_1, \omega_2, \omega_3 \dots ,  \omega_{N_{\omega}}$}
\State Calculate $W(\tau_i)$ using \eqref{exactFFT}\Comment{$N^2N_{\tau} N_{\omega}$}
\EndFor
 \For {$A \in N_{atom}, B \in N_{atom}$}
  \State Evaluate $\Sigma^{AB}(\tau_i)$ using \eqref{sigmaAll} to \eqref{sigmaIII} 
  \Comment{$N^2 N_{\tau}\phantom {N_{\omega}}$}
  \EndFor 
   \State Calculate $\Sigma_{nn,\tau_i}$ and evaluate \eqref{fft3} 
   \EndFor  
\State Evaluate QP-spectrum using \eqref{quasi-particle-equations}
\end{algorithmic}
\caption{Pseudocode for $G_0W_0$ using PADF. The asymptotic operation count of some key steps is given on the right.}
\label{alg::GW}
\end{algorithm}

\section{\label{sec::Results}Results}
\subsection{Computational Details}
All calculations have been performed with a locally modified development version of ADF\cite{Snijders2001, ADF2019} in which the herein described PADF-$G_0W_0$ algorithm has been implemented. In all gKS calculations, PADF has been used to evaluate Coulomb- and exchange terms.\cite{Guerra1998, Franchini2014,Krykunov2009} We performed PADF-$G_0W_0@$PBE and PADF-$G_0W_0@$PBE0 calculations for all molecules in the GW100 database\cite{VanSetten2015} as well as PADF-$G_0W_0@$PBE0\cite{Ernzerhof1999, Adamo1999} calculations for the 50 largest molecules in the GW5000 database.\cite{Stuke2020} For GW100, we used the structures as published in the original work\cite{VanSetten2015}, except for Vinylbromide and Phenol for which we used the updated structures.\cite{vanSetten2020} To preclude potential confusion, we emphasize that all results from other codes we refer to herein have been taken from the literature and have not been calculated by us. 

We herein use several all-electron (AE) STO-type basis sets of double-$\zeta$ (DZ), TZ, and QZ size. The prefix \enquote{aug-} denotes augmentation of a basis set with an additional shell of diffuse functions for all angular momenta $l =0, 1, 2$. For augmented QZ basis sets, an additional diffuse shell of f-functions is added as well. Augmentation of the basis set with $x$ additional shells of polarization functions is denoted by $x$P. We employ two different QZ basis sets, the even-tempered QZ3P\cite{Chong2004} basis set, and the larger QZ4P\cite{vanLenthe2003} basis set. For a detailed description of the basis sets we refer to van Lenthe et al..\cite{vanLenthe2003} It should be noted, however, that all basis sets are not correlation consistent (CC) and unsuitable for CBS limit extrapolation. Also note, that QZ3P and all augmented basis sets are only available for the first 4 rows of the periodic table. In case of QZ3P we will use QZ4P for all heavier elements and in case of augmented basis sets we use the respective basis set without augmentation. 

If not indicated otherwise, we used the \emph{Normal} auxiliary fit set,\bibnote{Details of the composition of the fit sets can be found in the supporting information of our recent work\cite{Forster2020}.} \emph{Good} quality for numerical integration\cite{ Franchini2014}, \emph{Normal} quality for thresholds controlling distance effects, and standard numerical settings otherwise. We use imaginary time and frequency grids with up to $N_{\omega} = $ 18 points each\bibnote{The individual numbers of points can differ. This is due to the fact that we adjust the grid sizes at runtime to match a certain error parameter. Thus, for many systems, the number of points will actually be smaller than 18, since the imaginary time and frequency integrals are already converged with a smaller number of points.} for GW100 and $N_{\omega} = $ 16 points each for GW5000 and use a padé-approximant of order $N_{\omega}$ to model the self-energy on the real frequency axis. In all $G_0W_0@PBE0$ calculations on GW5000 we employed the unscaled Zero Order Regular Approximation (ZORA).\cite{VanLenthe1993, VanLenthe1994, VanLenthe1996, VanLenthe1999}

During orthonormalization of the Fock matrix in the SCF, columns of the transformation matrix are removed when the corresponding eigenvalues of the AO-overlap matrix are smaller than some threshold $\epsilon_D$.\cite{Lowdin1967} As explained above, we have adjusted this value to $\epsilon_D = 10^{-3}$ in all calculations using QZ4P or augmented basis sets. Otherwise, the default of $\epsilon_D = 10^{-4}$ has been used. 

\subsection{Benchmarks}

\subsubsection{The GW100 database}

The size and type of basis set is the most crucial factor influencing the results of a GW calculation.\cite{VanSetten2015} Using localized basis functions, even on the QZ level BSEs for HOMO and especially LUMO QP energies can exceed several hundreds meV, necessitating an CBS limit extrapolation to obtain very accurate reference values.\cite{VanSetten2015, Golze2019, Stuke2020} Using localized AOs one needs to rely on heuristics since the expansion of MOs in this basis does not converge uniformly, unlike expansions in terms of PW\cite{Payne1992} or finite elements in RS.\cite{Varga2004} HOMO QP energies obtained with these basis set types are generally in good agreement with the original ones by van Setten et al.,\cite{VanSetten2015} while differences for unbound LUMO energies often exceed 1 eV.\cite{Maggio2017a, Govoni2018, Gao2019} 

\begin{figure}[h]
    \centering
    \includegraphics[width=\textwidth]{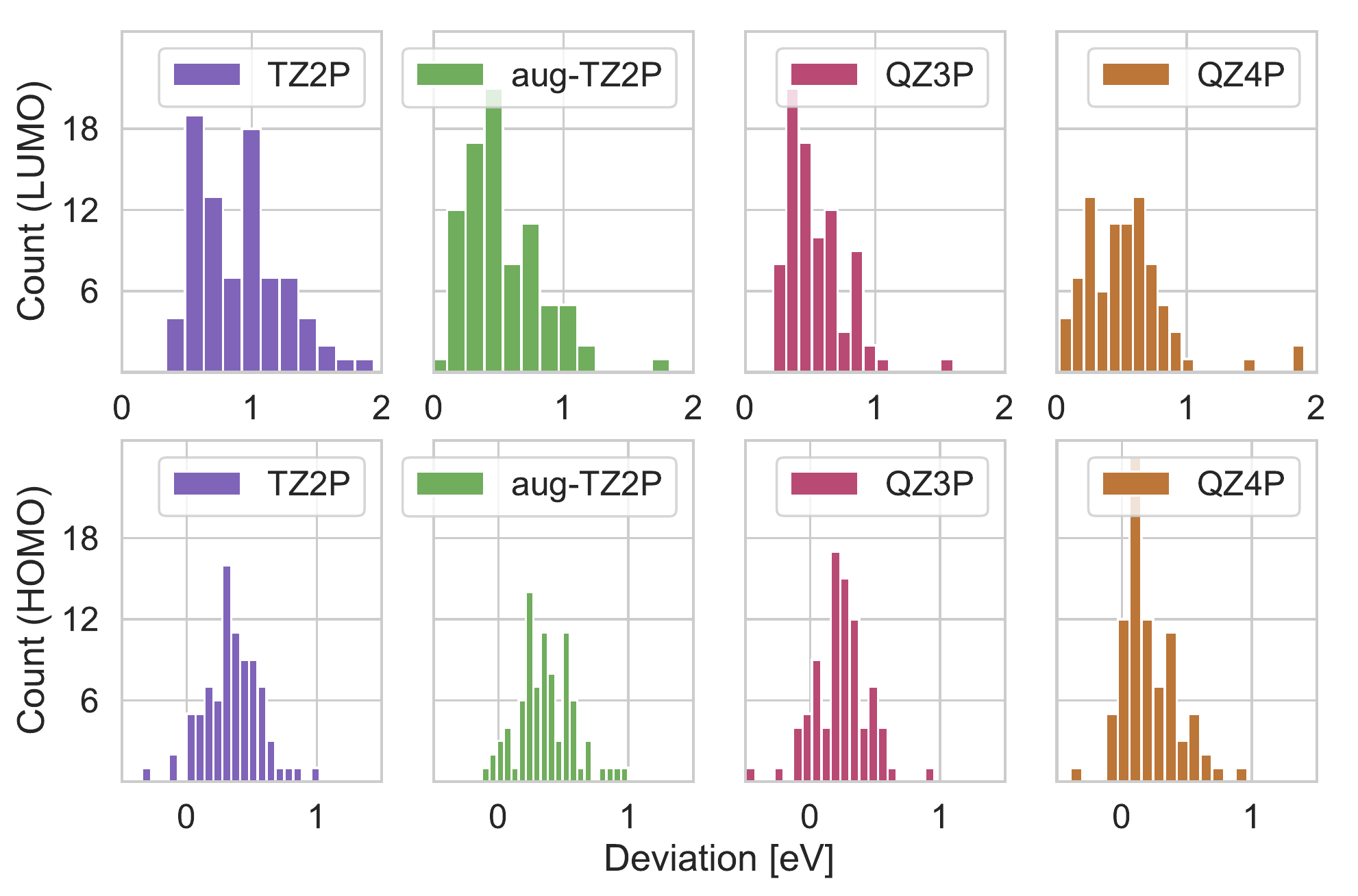}
    \caption{Error distributions (in eV) for $G_0W_0@$PBE with four different STO-type basis sets for the HOMO (bottom) and LUMO QP energies (top) in the GW100 database with respect to the nanoGW reference. Due to its error larger than 3 eV, \ce{CO2} is not included in the upper left plot.}
    \label{fig::gw100_HOMO}
\end{figure}

Thus, it is not straightforward for our purpose to chose appropriate reference values and we will therefore use more than one reference in the following. As primary reference for GW100 we use RS results\bibnote{These reference values have been calculated by Chelikowsky and coworkers with the nanoGW\cite{Tiago2006} package which implements $GW$ in RS and with a full frequency treatment. The calculations have been performed using KS orbitals and energies calculated with the PARSEC code.\cite{Chelikowsky1994, Kronik2006} For details we refer to the original work\cite{Gao2019}} by Chelikowsky and coworkers.\cite{Gao2019} While these results are not extrapolated to the CBS limit, they are carefully converged and should be a very reliable reference. This choice is mainly motivated by the large differences between plane-wave and GTO implementations for unbound LUMO energies. It has been argued by Kresse and coworkers\cite{Maggio2017a} that the GTO-type basis sets of the def2 family might not be flexible enough to adequately describe these QP energies and lead to significantly overestimated unbound LUMO energies. As also pointed out in ref. \citen{Maggio2017a}, CC GTO-type basis sets are much more suitable in this respect, however, since we are not aware of reference values for GW100 this is not an option for benchmarking. On the other hand, the nanoGW results deviate to WEST/VASP by only 134/122 meV for GW100, only excluding all noble gases and \ce{H2} but including all other molecules with unbound LUMOs.\cite{Gao2019} While in comparisons between different codes these systems are often excluded,\cite{Govoni2018} we decided to retain them in this work as well. However, we excluded from our analysis all noble gases and \ce{H2} since for these molecules the discrepancies between RS, PW-pseudopotential and AE codes often exceed 2 eV.\cite{VanSetten2015, Maggio2017a, Govoni2018, Gao2019} We also excluded \ce{CI4}, \ce{KBr}, \ce{NaCl}, \ce{BN}, \ce{O3}, \ce{BeO}, \ce{MgO}, \ce{Cu2} and \ce{CuCN} for which multiple solutions can be found when the QP equation \eqref{quasi-particle-equations} is solved for the HOMO. This leaves us with a set of 85 molecules which we will discuss in the following. In section \ref{sec::gw5000}, we benchmark our implementation against reference results obtained with GTOs for a large number of organic molecules with bound LUMOs.

\begin{figure}[h]
    \centering
    \includegraphics[width=0.8\textwidth]{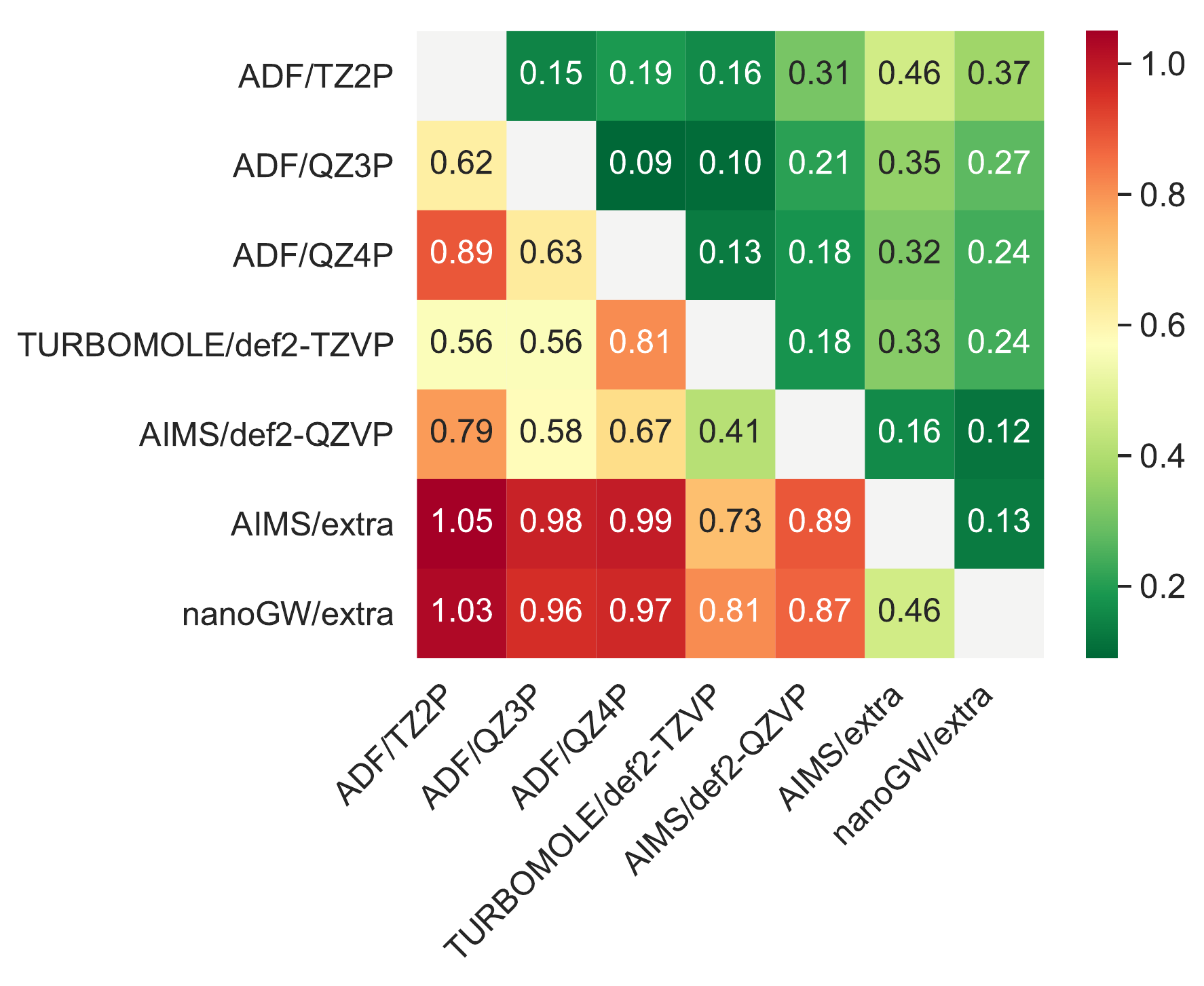}
    \caption{MADs (upper triangle) and maximum absolute deviations (lower triangle) in eV of QP HOMO energies computed with different codes and basis sets, specified on the axes for $G_0W_0@$PBE.\cite{vanSetten2020}}
    \label{fig::gw100cross}
\end{figure}

The histograms in figure~\ref{fig::gw100_HOMO} summarize the results of our benchmarks on GW100 and shows errors obtain with our implementation and different basis sets with respect to the nanoGW reference. For individual QP energies we refer to the supporting information. Figure~\ref{fig::gw100cross} shows MADs for the HOMO QP energies between different codes and basis sets. Since we are not able to perform basis set extrapolation, the QZ4P results are the best ones attainable for us. We observe MADs of 0.24 eV with respect to the nanoGW\cite{Tiago2006} results and of 0.32 eV to the CBS limit extrapolated (CBSLE) FHI-AIMS\cite{FHIaims2009,Blum2009, Ren2012} QP energies. With respect to both, the RS and the FHI-AIMS CBS limit, QZ4P yields an accuracy comparable to TURBOMOLE\cite{Balasubramani2020} with the smaller def2-TZVP basis set (6th and 7th column in the heatmap in figure~\ref{fig::gw100cross}). QZ4P does not give significant improvements over QZ3P and with a MAD of 0.12 eV with respect to the nanoGW reference, def2-QZVP performs considerably better than QZ4P. The fact that the former one has more polarization functions (e.g. (7s,4p,3d,2f,1g) vs. (7s,4p,2d,2f)  for second row elements\cite{Pritchard2019}) might explain part of the discrepancy. Furthermore, the excessive truncation of the QZ4P-basis in the canonical orthonormalization procedure during the SCF effectively diminishes the size of the virtual space. This might also explain why QZ4P only improves moderately over the significantly smaller TZ2P basis set (0.46 vs. 0.32 eV) while going from def2-TZVP to def2-QZVP reduces the MAD with respect to both CBS limits by roughly 50 \%. Also a visual inspection of the error distributions for the QZ3P and QZ4P QP HOMO energies in figure~\ref{fig::gw100_HOMO} reveals that the QZ4P-errors shows a larger spread and more often exceed 0.5 eV than for QZ3P.

\begin{table}
    \centering
    \begin{tabular}{lcccc}
    \toprule
     & TZ2P & aug-TZ2P & QZ3P & QZ4P \\
    \midrule
    HOMO & 0.37 & 0.37 & 0.27 & 0.24 \\
    LUMO & 0.94 & 0.55 & 0.56 & 0.52 \\
    gap  & 0.59 & 0.26 & 0.34 & 0.35 \\
    \bottomrule
    \end{tabular}
    \caption{MADs of the $G_0W_0@$PBE HOMO and LUMO QP energies corresponding to figure~\ref{fig::gw100_HOMO} and HOMO-LUMO gaps with respect to the CBS limit for four different STO-type basis sets (All values in eV).}
    \label{tab::mads}
\end{table}

For the LUMO QP energies shown in the upper part of figure~\ref{fig::gw100_HOMO}, aug-TZ2P, QZ3P and QZ4P show comparable MADs of 0.55, 0.56 and 0.52 eV, respectively and thus improve significantly over TZ2P with a MAD of nearly 1 eV (see table~\ref{tab::mads}). For TZ2P and both QZ basis sets, the MAD of the LUMO QP energies with respect to the RS reference values are more than twice as large than for the HOMO which results in a rather poor description of the HOMO-LUMO gap. This behaviour is similar to the performance of the def2 family of GTO-type basis sets for GW100 for which the errors for the LUMO are on average roughly twice as large than for the HOMO.\cite{VanSetten2015} aug-TZ2P overestimates HOMO and LUMO QP energies most symmetrically and consequently, with a MAD of 0.26 eV, describes the HOMO-LUMO gap significantly better than both QZ basis. The situation is well known from augmented GTO-type basis sets\cite{Blase2011, Boulanger2014, Faber2015, Jacquemin2015, Bruneval2015,Wilhelm2016a} which usually converge considerably faster to the CBS HOMO-LUMO gap than non-augmented basis sets,\cite{Wilhelm2016a} although the individual HOMO and LUMO levels are often not converged at all. 

\begin{figure}
    \centering
    \includegraphics[width=\textwidth]{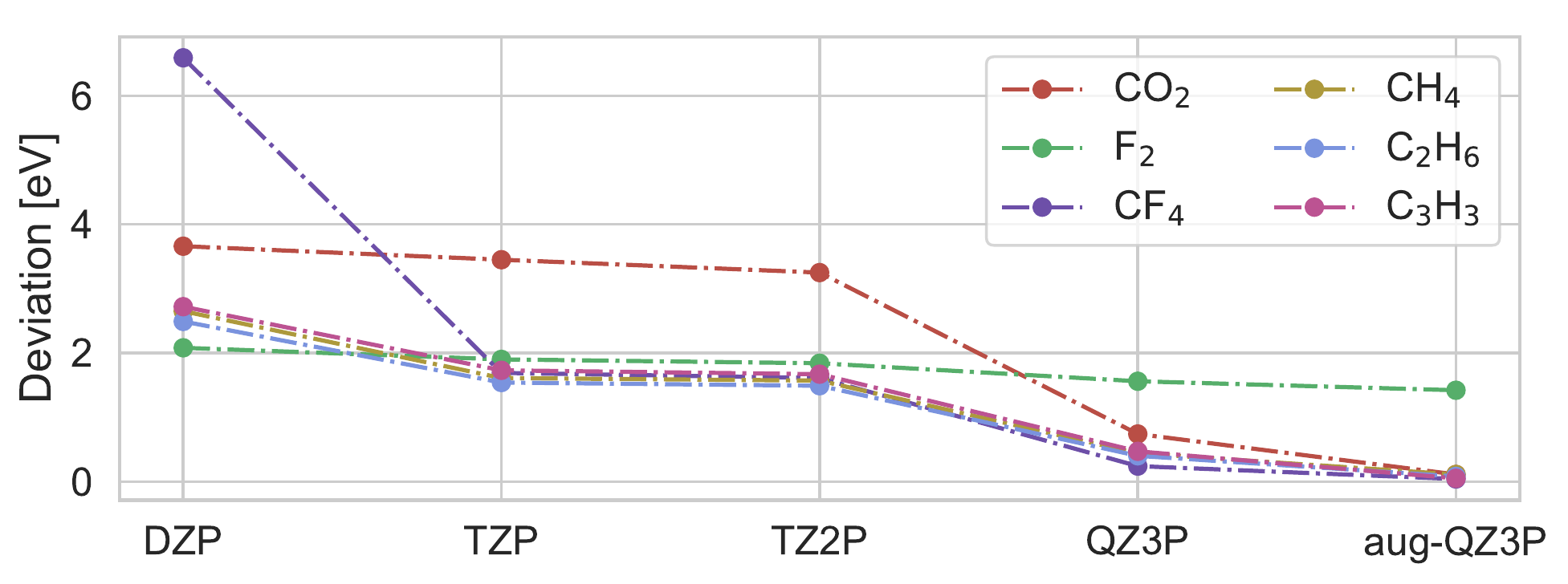}
    \caption{Deviations of $G_0W_0@$PBE LUMO QP energies to the CBS limit for six selected molecules from the GW100 database for different STO-type basis sets (all values in eV).}
    \label{fig::bse_convergence}
\end{figure}

Finally, we investigate some LUMO QP energies with exceptionally slow convergence to the CBS limit in more detail and see whether convergence can be attained using larger basis sets, keeping in mind the restrictions imposed by the PADF as explained in section~\ref{sec::theory}; reaching the basis set limit is only possibly for us for very small systems. While the \ce{CO2} LUMO QP energy deviates from the CBS limit by more than 3 eV, also for \ce{F2}, \ce{CF4} \ce{C3H3} and $\text{C}_n\text{H}_{2n+2}$ for $n=1, \dots 4$, the TZ2P LUMO QP energies deviate between 1.7 eV and 1.4 eV from the CBS limit. We investigate the convergence with respect to the basis set size for these molecules (except for Propane and Butane) in figure~\ref{fig::bse_convergence} by adding diffuse functions to the QZ3P basis set. 

{}

For all molecules except \ce{F2}, our aug-QZ3P results agree very well with the RS reference values. This is a little surprising since the GTO-type basis set CBSLE results differ by more than 1 eV for \ce{CH4}, \ce{C2H6} and \ce{C3H3}. In fact, not only for these systems we observe that our unbound LUMO energies are generally closer to RS and PW than to GTO references. While being out of the scope if this work, this is an interesting observation which deserves further investigation. For \ce{F2}, the extrapolated CBS limits from different codes are in good agreement and the errors from aug-QZ3P are still hard to explain with BSEs alone, although van Setten et al. found a BSE of 0.53 eV for the LUMO energy using def2-QZVP.\cite{VanSetten2015}

\subsubsection{\label{sec::gw5000}The GW5000 database}
We now turn our attention to systems large enough for local approximations to take effect and discuss the HOMO and LUMO energies of 20 organic molecules with in between 85 and 99 atoms from the GW5000 database.\cite{Stuke2020} These tests are crucial for our purpose. First, they allow us to assess the effect of the values of the thresholds controlling distance effects. As explained in detail elsewhere\cite{Forster2020}, we essentially rely on three thresholds in our implementation, which we organize in three tiers, denoted as \emph{Basic}, \emph{Normal} and \emph{Good}. For the exact values of these thresholds we refer to the supporting information.

\begin{figure}
    \centering
    \includegraphics[width=\textwidth]{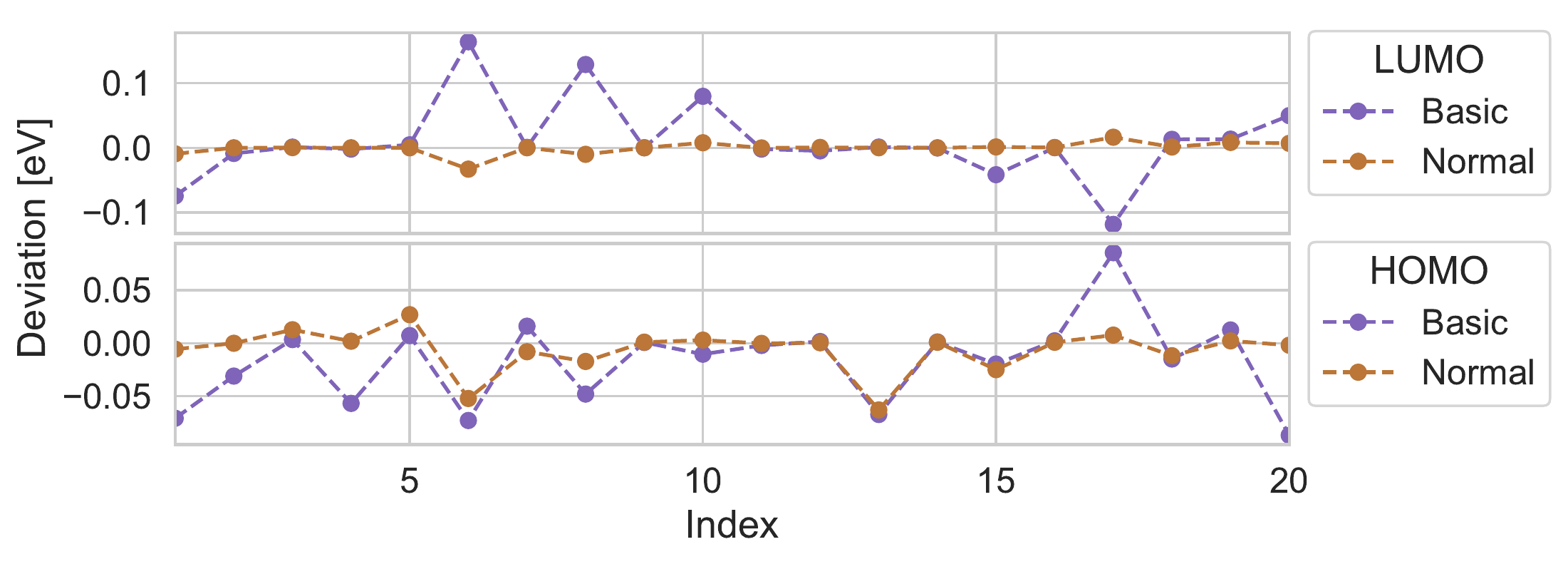}
    \caption{Deviations of the \emph{Basic} and \emph{Normal} threshold tiers with respect to the \emph{Good} tier for HOMO (bottom) and LUMO (top) QP energies on the $G_0W_0$/PBE0 level of theory (all values in eV).}
    \label{fig::gw5000_thresholds}
\end{figure}

The convergence with respect to the threshold tiers for HOMO and LUMO QP energies is shown in figure~\ref{fig::gw5000_thresholds}. As shown in the lower panel, the HOMO energies from different threshold tiers agree within 0.1 eV and the HOMO energies from the \emph{Normal} and the \emph{Good} threshold tier usually agree within an accuracy of 60 mEV. Using the \emph{Basic} threshold tier, the LUMO QP energies show a maximum deviation of roughly 0.15 meV with respect to the \emph{Good} tier. On the other hand, the LUMO energies from the \emph{Normal} and \emph{Good} tier are in even better agreement than the corresponding HOMO energies. Thus, using the \emph{Normal} tier ensures an internal precision of our implementation of 60 meV for HOMO and LUMO QP energies. In case only the HOMO level is of interest, sufficient precision is already attained using the \emph{Basic} threshold tier. 

\begin{figure}[h]
    \centering
    \includegraphics[width=\textwidth]{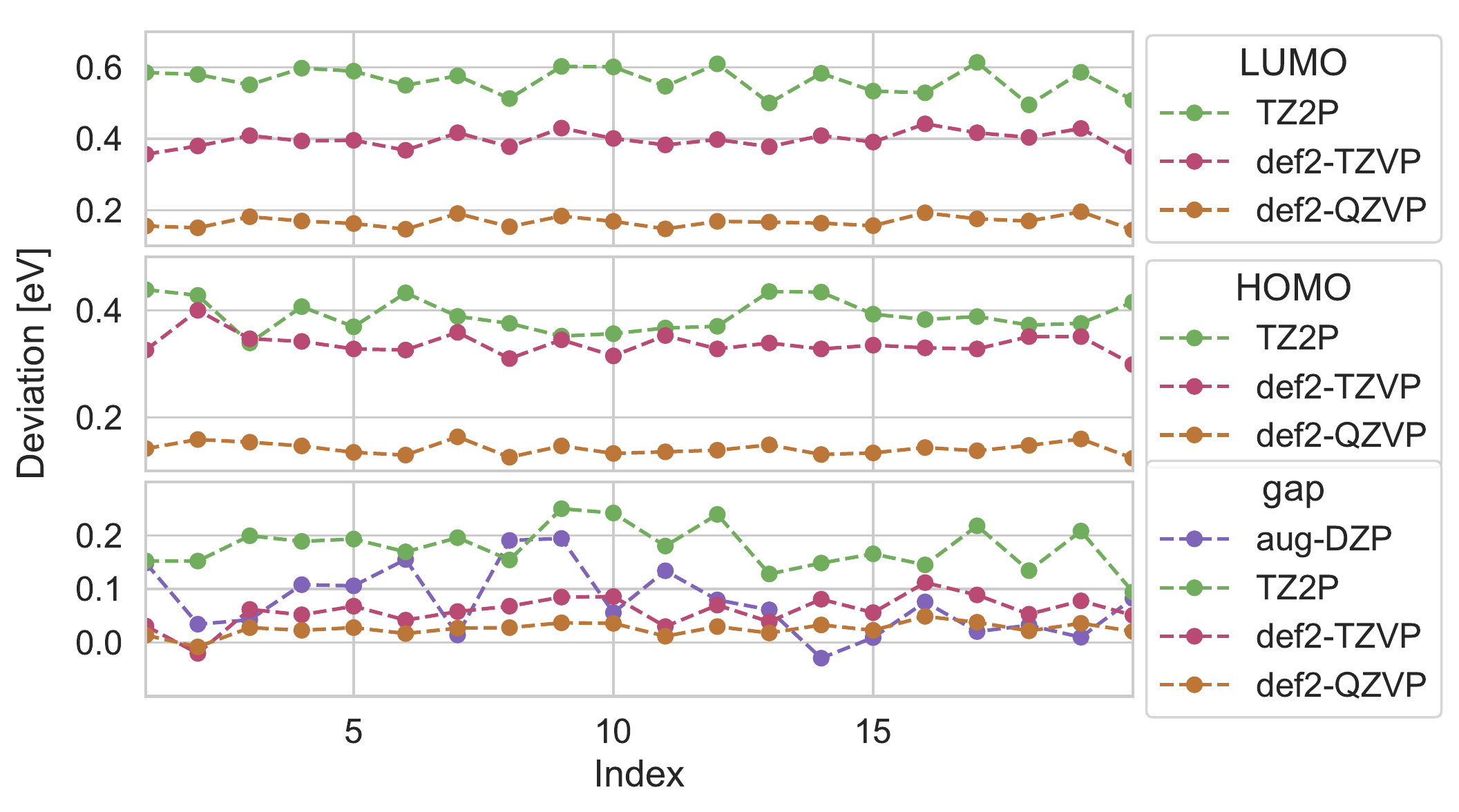}
    \caption{Deviations of LUMO (upper panel) QP energies, HOMO (middle panel) QP energies and HOMO-LUMO QP gaps (lower panel) for the TZ2P (\emph{Normal} thresholds) as well as the GTO-type def2-TZVP and def2-QZVP basis sets with respect to the CBS limit for the HOMO energies of the 20 large molecules from the GW5000 database. HOMO-LUMO QP gaps from aug-DZP are shown as well. All values are in eV).}
    \label{fig::gw5000}
\end{figure}

Second, the applicability of our implementation to the small molecules in GW100 does not imply the same for larger systems. In fact, this is true for any method exploiting locality in any form. Due to the reasons outlined in section~\ref{sec::theory} we refrain from reporting results with QZ and large augmented basis sets for these systems. Instead, we want to investigate the accuracy attainable using the TZ2P and aug-DZP basis sets for which no numerical problems can be expected also for large molecules.

\begin{table}[]
    \centering
    \begin{tabular}{lcccc}
    \toprule
              & TZVP & QZVP & TZ2P & aug-DZP \\
    \midrule
         HOMO &0.34 & 0.14 & 0.39 & 0.46  \\
         LUMO &0.40 & 0.17 & 0.56 & 0.53  \\
         gap  &0.06 & 0.03 & 0.18 & 0.08  \\
    \bottomrule
    \end{tabular}
    \caption{MADs of HOMO energies, LUMO energies and HOMO-LUMO gaps with respect to the CBS limit for the 20 considered molecules from the GW5000 database for different basis sets. The \emph{Normal} tier of thresholds has been used in all PADF-$G_0W_0$ calculations. All values are in eV. }
    \label{tab::gw5000_mads}
\end{table}

We compare our results for QP HOMO and LUMO levels as well as HOMO-LUMO gap for the 20 selected molecules to accurate reference values calculated with numerical GTOs with the FHI-AIMS code in figure~\ref{fig::gw5000}. MADs for these quantities with respect to the CBS limit are given in table~\ref{tab::gw5000_mads}. We observe that the TZ2P HOMO QP energy never deviates from def2-TZVP by more than 0.1 eV and the MAD of 0.39 eV is only 50 meV higher than the one found for def2-TZVP. For the LUMO energy, the situation is different. While def2-TZVP yields a MAD of 0.40 eV for this quantity, TZ2P performs with 0.56 eV considerably worse. This has a profound effect on the description of the HOMO-LUMO gap. Since def2-TZVP overestimates the LUMO level not much more than the HOMO QP energy, the HOMO-LUMO gap shows with a MAD of 0.06 eV an excellent agreement to the CBS limit while TZ2P yields a MAD of 0.17 eV. On the other hand, using the smaller aug-DZP basis set we find with a MAD of 0.08 eV good agreement with the CBS limit. As might be inferred from table~\ref{tab::gw5000_mads}, this success results mainly in poorer description of the HOMO level compared to TZ2P and the error cancellation between HOMO and LUMO is not always reliable, which can be seen from systems $\#$8 and $\#$9 whose HOMO-LUMO gap differs to the CBS limit by 0.2 eV. It should also be noted, that aug-DZP calculations are slightly slower than TZ2P ones for medium and large systems since more AO-pair products need to be considered.

\begin{figure}[h]
    \centering
    \includegraphics[width=\textwidth]{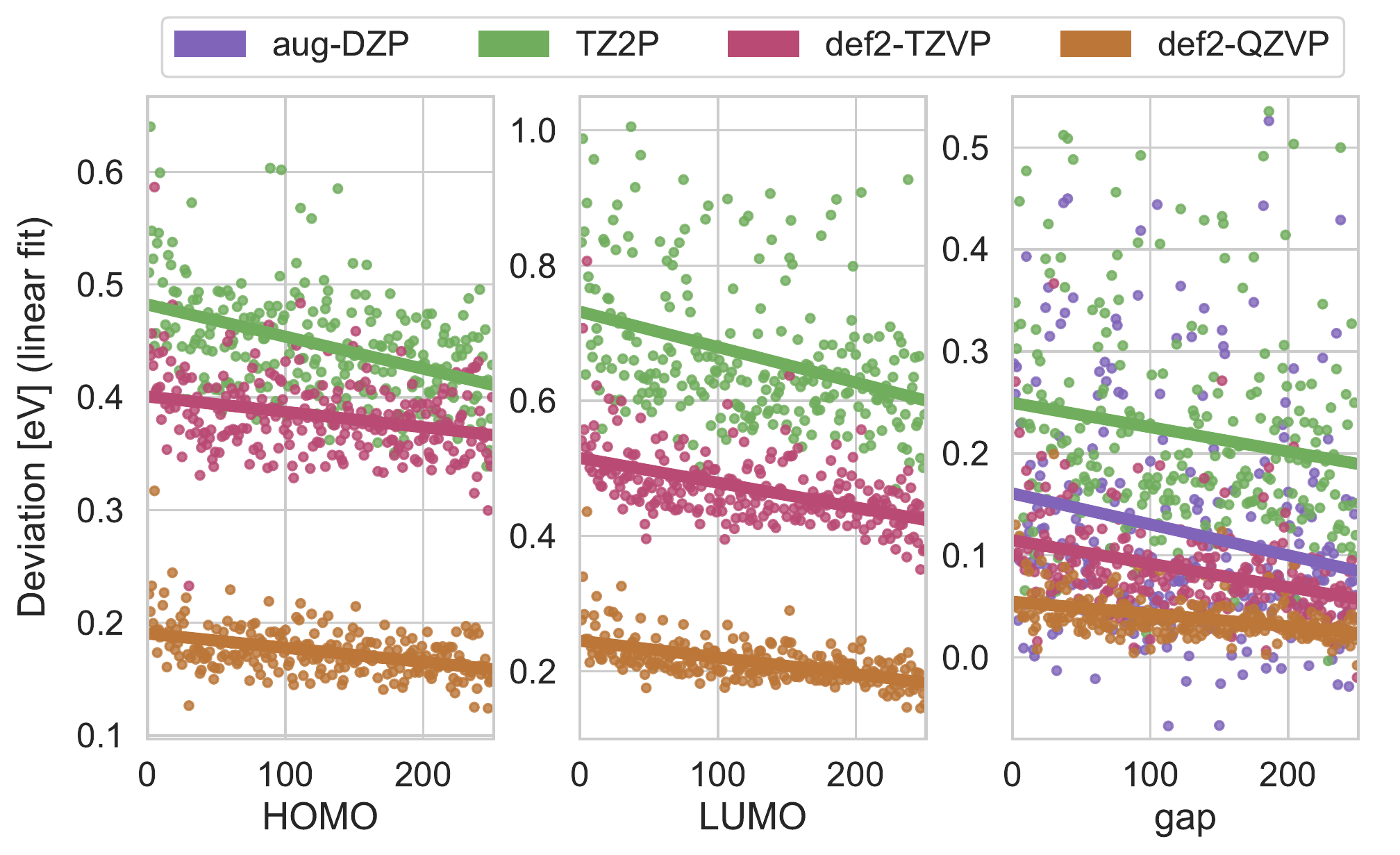}
    \caption{HOMO (left), and LUMO QP energies (middle) as well as HOMO-LUMO QP gaps (right) with different basis sets for 250 randomly selected molecules from the GW5000 database (dots) as well as linear fits, $f(x) = a \times x + b$. The systems have been sorted according to increasing size.}
    \label{fig::s250}
\end{figure}

Finally, we investigate the accuracy of our algorithm as a function of systems size. To this end, we randomly selected 250 molecules from the GW5000 database and sorted these systems from smallest (12 atoms) to largest (99 atoms). Figure~\ref{fig::s250} shows the deviations to the CBS limit of our $G_0W_0@$PBE0 results for HOMO, LUMO and HOMO-LUMO QP gap with the TZ2P and aug-DZP basis sets as well as FHI-AIMS results using the def2-TZVP and def2-QZVP basis sets\cite{Stuke2020}. Additionally, we performed linear fits as implemented in Numpy\cite{Harris2020}, which are also shown in figure~\ref{fig::s250}. Essentially we obtain the same picture as for the 20 large molecules: TZ2P performs nearly as good as def2-TZVP for the HOMO QP energies and considerably worse for the LUMO level which translates into a worse description of the HOMO-LUMO gap. While it is observed that the STO-results show a larger spread than their GTO counterparts especially for LUMO energies, we also observe that the deviation to the CBS limit decreases with growing system size for all basis sets. For all subplots in figure~\ref{fig::s250}, the TZ2P fit is more or less parallel (also see the fit-parameters in the supporting information for comparison) to the GTO-fits, while the slope in the aug-DZP fit for the HOMO-LUMO gap is slightly more negative. As for the subset of 20 large molecules, aug-DZP produces HOMO-LUMO gaps which on average agree with the CBSLE reference within 0.15 eV for systems larger than a few tens of atoms. However, in some cases the errors can still be rather large (e.g. larger than 0.4 eV in 7 out of 250 cases), while the def2-QZVP BSE practically never exceeds 0.1 eV.

The decreasing errors are most likely due to basis set superposition which leads to a more complete basis when the system increases and the assumption that this effect is more pronounced for basis sets with many diffuse functions such as aug-DZP is reasonable. Thus, we can conclude that the accuracy of our algorithm is not negatively affected by the system size. We note, that local over-completeness and the associated numerical issues can already be encountered for very small systems like the ones the left side of the plots in figure~\ref{fig::s250}. On the other hand, it is highly unlikely that they become more pronounced for larger systems due to the locality of the AOs.

\subsection{Representative timings}

\begin{figure}
    \centering
    \includegraphics[width = \textwidth]{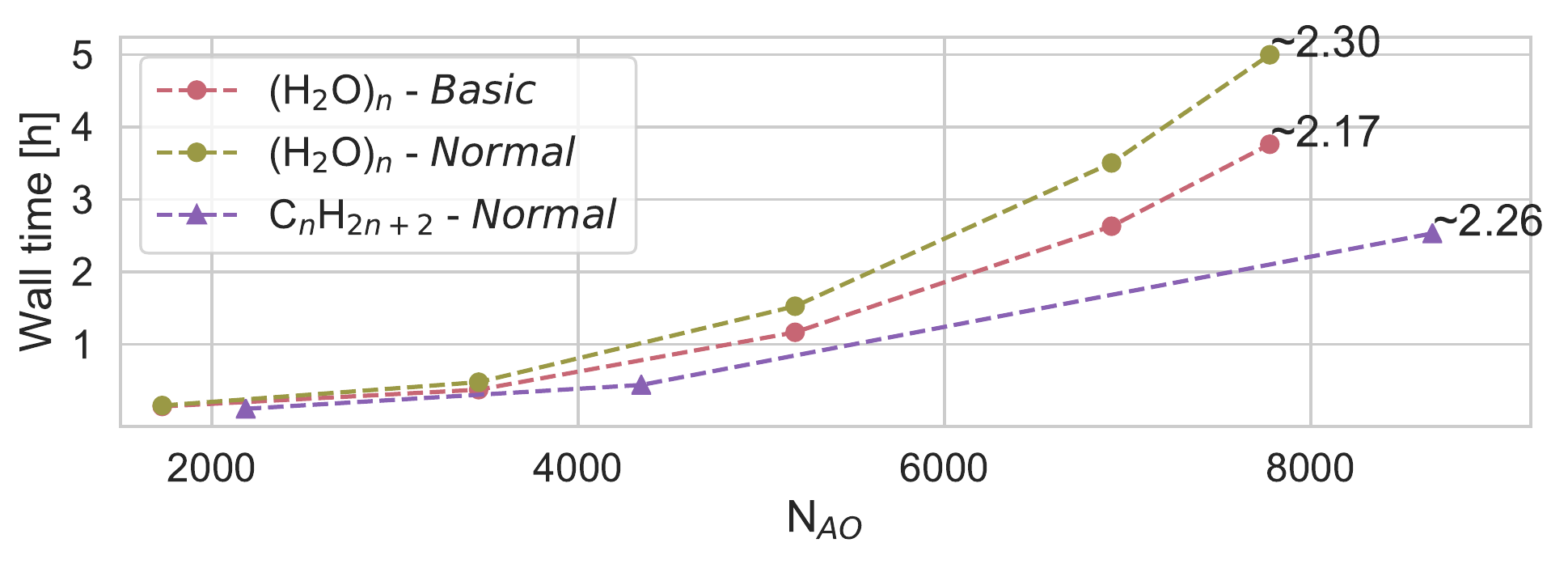}
    \caption{Wall times in hours for $G_0W_0$@PBE/TZ2P calculations on a series of Water clusters and a linear alkane chain (exclusive the preceding SCF). All calculations have been performed on 2 bw nodes. The exponent of the polynomial describing the asymptotic scaling of the algorithm is given on the right of each plot.}
    \label{fig::timings}
\end{figure}

\begin{figure}
    \centering
    \includegraphics[width=\textwidth]{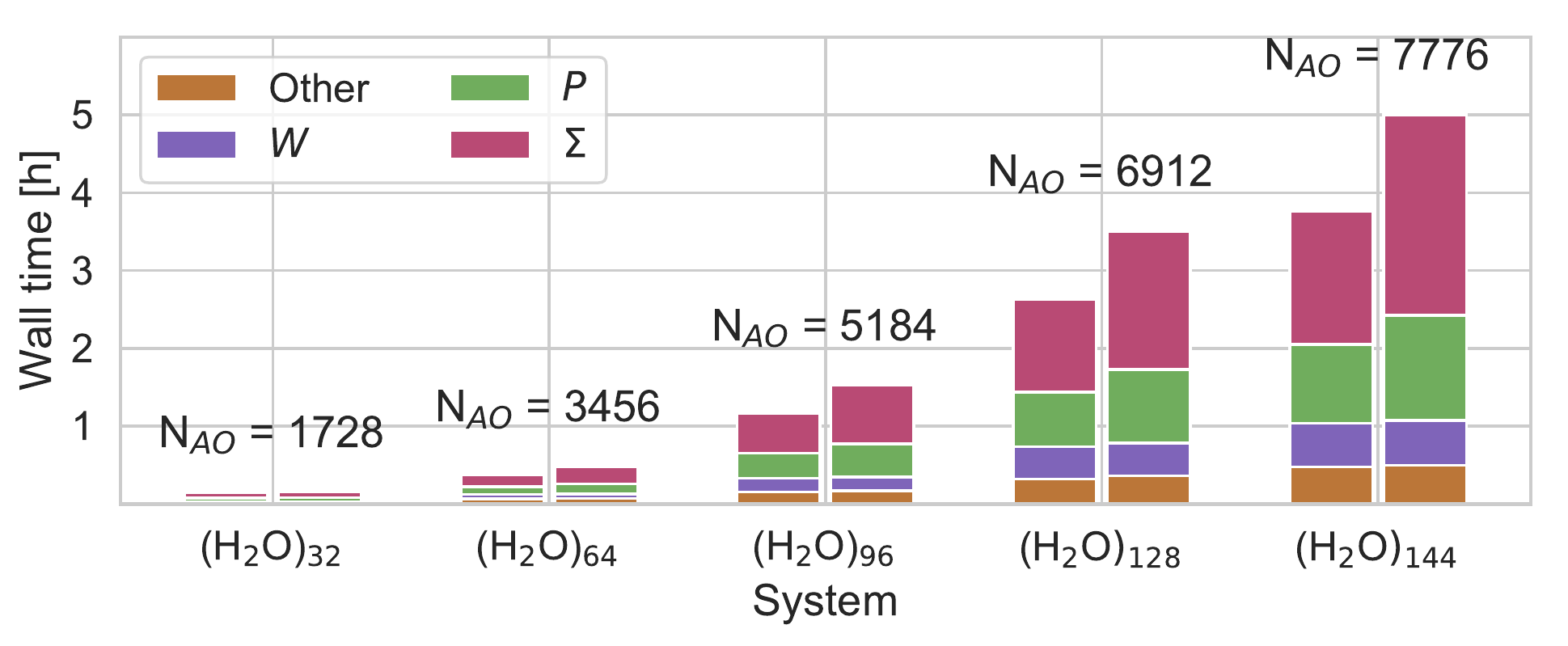}
    \caption{Contributions to total $G_0W_0$ wall times from different key steps for a series of Water cluster using the TZ2P basis set. Left bar in each group: \emph{Basic} threshold quality, right bar in each group: \emph{Normal} threshold quality. All calculations have been performed on 2 bw nodes.}
    \label{fig::timings_details}
\end{figure}

In order to analyse the asymptotic scaling of our algorithm, we present $G_0W_0 @ PBE$/TZ2P calculations on series of water clusters\bibnote{The structures of the water clusters have been downloaded from the website of the ERGO program,\cite{Rudberg2018} \href{http://www.ergoscf.org}{http://www.ergoscf.org} (visited on may 19th, 2020).} using the same numerical settings as for GW5000, the \emph{Basic} and \emph{Normal} tiers of thresholds and 12 imaginary time and imaginary frequency points. All calculations presented in this subsection were performed on 2.2 GHz intel Xeon (E5-2650~v4) nodes (broadwell architecture) with 24 cores and 128 GB RAM each (bw nodes in short). Figure~\ref{fig::timings} shows the wall times for the $G_0W_0$-part of the calculations and the exponents of the polynomials describing the asymptotic scaling of these calculations with increasing system size. Information on CPU time and asymptotic scaling of key steps of the algorithm for the largest of these systems are given in figure~\ref{fig::timings_details}

The largest water cluster here comprises 432 atoms with 7776 AOs and 36576 ABFs. Using the \emph{Normal} threshold tier, the whole $G_0W_0$ calculation takes five hours on two nodes. As shown in figure~\ref{fig::timings_details}, the most expensive step is the calculation of $\Sigma$, being responsible for about half of the wall time of the whole calculation, followed by the evaluation of $P$. The evaluation of $\Sigma$ is also the step which is accelerated most when the thresholds are loosened. This is due to the contractions eq. \eqref{intermediateI} which are tremendously accelerated when the multipole approximation is used for an increasing number of atom pairs. Consequently, the asymptotic scaling of this step is decreased from $N^{2.34}$ to $N^{2.15}$. Also the asymptotic scaling of $P$ is reduced considerably (from $N^{2.19}$ to $N^{2.05}$), so that the wall time of the total calculation can be reduced to less than 4 hours. Note, that the evaluation of $W$ is not affected by changing the thresholds and asymptotically scales as $N^3$. However, even for the largest water cluster the timings are clearly dominated by $P$ and $\Sigma$ and $W$ can not be expected to become a bottleneck even for systems much larger than the ones considered here.

Water clusters are very compact systems due to their spherical shapes. This takes an adverse effect on the asymptotic scaling properties of our algorithm, compared to low-dimensional systems, e.g. linear alkane chains as the most extreme example. The timings for a series of alkane chains is given for comparison in figure~\ref{fig::timings} as well. With the same thresholds, the $G_0W_0$-calculation for \ce{C160H322} takes with roughly 2.5 hours only half the time as the one for \ce{(H2O)144} even though the former system is larger. In fact, $P$ is calculated in less than half an hour which is less wall time than is required for the calculation of $\widetilde{W}$. 

\begin{figure}[h]
    \centering
    \includegraphics[width=\textwidth]{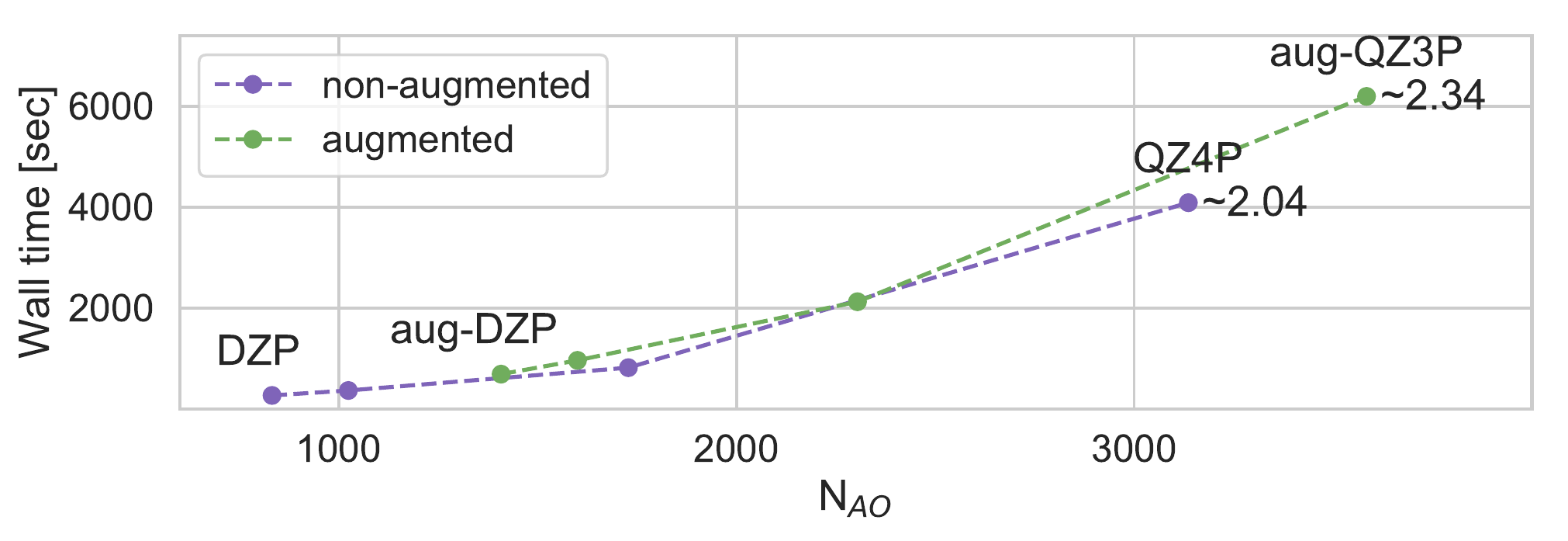}
    \caption{Timings (in seconds) and asymptotic scaling of our algorithm with basis sets of increasing size for \ce{(H2O)32} with the same settings as described above (\emph{Normal} thresholds) using a single bw node.}
    \label{fig::scaling_basis}
\end{figure}

Next, we investigate scaling with respect to the single-particle basis at fixed systems size. As shown in figure \ref{fig::scaling_basis}, even for the small \ce{(H2O)32} cluster, our algorithm scales quadratic with the size of the single-particle basis when non-augmented basis sets are used. Using augmented basis sets, the asymptotic scaling is worse owing to the large number of basis functions with a very slow decay with the distance to the nuclei on which they are centred, leading to a smaller number of negligible AO-products. High scaling with respect to the single-particle basis is a general shortcoming of AO-based algorithms compared to MO-based ones and its current form it is difficult to envisage modifications of our algorithm which might overcome this issue.

\begin{figure}[h]
    \centering
    \includegraphics[width=\textwidth]{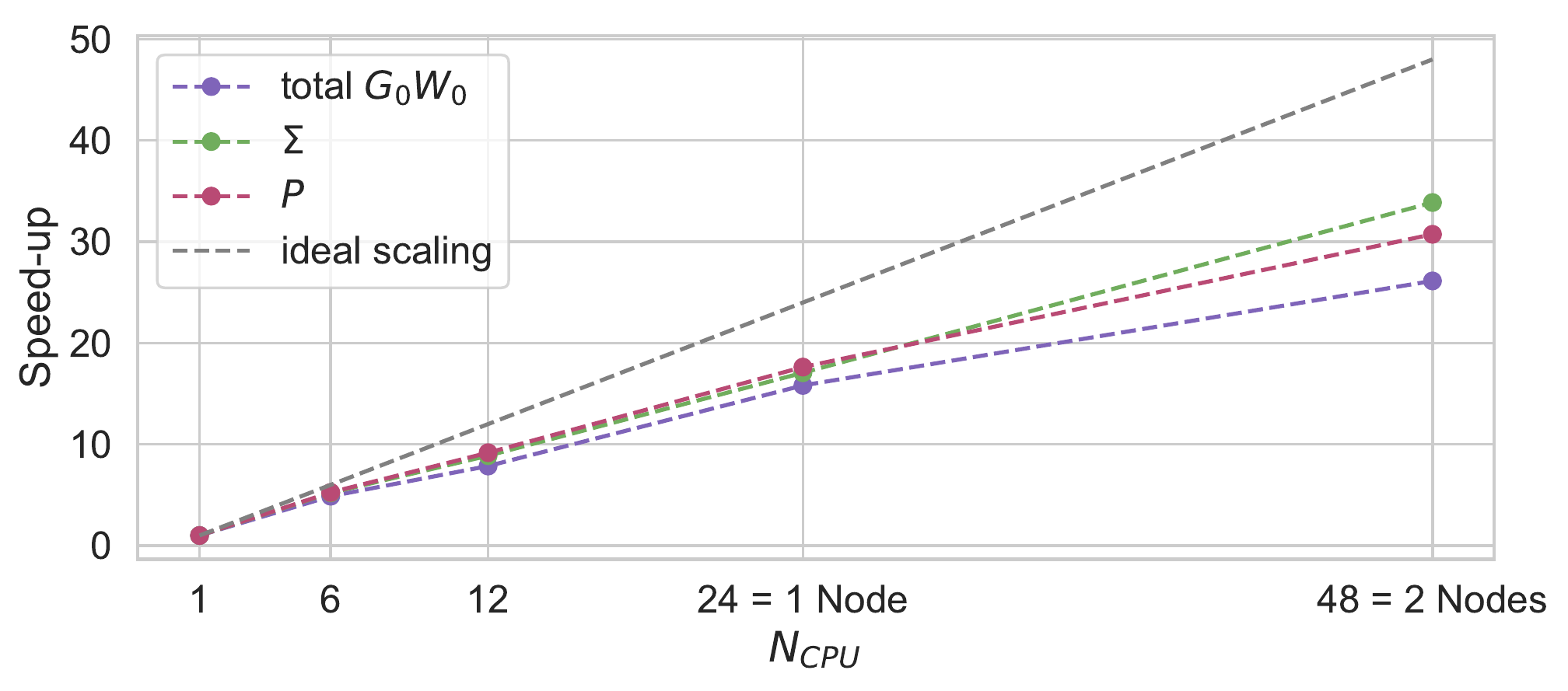}
    \caption{Speed-up for \ce{(H2O)64} (same settings as described above, \emph{Normal} thresholds) with the number of CPUs.}
    \label{fig::scaling_nodes}
\end{figure}

Finally, we comment on the parallel performance of our algorithm. Figure~\ref{fig::scaling_nodes} shows the speed-up with increasing number of cores.
We achieve a parallel efficiency of 66 \% when going from 1 to 24 cores. The deviation to the ideal speed-up is partly due to small fractions of serial code in our algorithm but also due to unnecessary network communication. Also due to the latter factor,  parallelization over multiple nodes is less efficient. At the moment, our algorithms for the calculation of $P$ and $\Sigma$ communicate a lot of data, an aspect which we have not optimized yet.

For completeness, we also mention the large memory count of our algorithm increasing as $N^2$ with systems size. However, the practical memory bottleneck is rather the storage of $C$. Although only linearly scaling, we store it in shared memory which prevents the scalability of our algorithm to even larger systems. The memory requirements are reduced for low-dimensional systems for which $C$ becomes smaller, however, it is clear that systems much larger than the ones presented herein can not be treated any more. Still, for systems of hundreds of atoms for which conventional implementations require a supercomputer,\cite{Wilhelm2018,DelBen2019}  $G_0W_0$ calculations with our algorithm can be performed in a routine fashion which puts its application in main-stream computational spectroscopy within reach.

\section{\label{sec::conclusion}Conclusion}

In this work, we have presented a PADF-based $G_0W_0$ implementation using STOs and relying on imaginary time-representation of the single-particle Green's function. Our algorithm combines quadratic scaling in memory and operation count with a very small prefactor due to a sparse map from ABF space to AO-product space. Using realistic numerical settings, a $G_0W_0$ calculation for a spherically shaped water cluster with 432 atoms, 7776 AOs, and 36576 ABFs takes 240 CPU hours. Using slightly looser thresholds, the same calculation is done in 180 CPU hours and the $G_0W_0$ calculations for a linear alkane chain with the same number of AOs takes only about 100 CPU hours. Thus, our algorithm is at least one order of magnitude faster than the fastest state-of-the-art canonical implementations.\cite{Wilhelm2016a, Wilhelm2018}

The accuracy of our algorithm for the calculation of the HOMO and LUMO QP energies in the GW100 database has been investigated by comparison to RS-CBSLE reference values. We found MADs of 0.38 eV for TZ2P and 0.26 eV for QZ4P for the HOMO, and 0.93 eV for TZ2P and 0.55 eV for QZ4P for the LUMO energies, respectively. For the HOMO level, FHI-AIMS/def2-QZVP only deviates from the CBS limit by 0.15 eV on average and TURBOMOLE/TZVP by 0.28 eV. Thus, for GW100, the accuracy of our algorithm on the QZ level is comparable to canonical implementations on the TZ level while it is difficult to make a definite statement about the quality of our LUMO energies due to large discrepancies between different codes.\cite{VanSetten2015, Maggio2017a, Govoni2018, Gao2019}

Two factors contribute to the relatively poor performance of our algorithm for GW100. First, for many systems with QP solutions close to poles of the self-energy, our frequency treatment with AC is inaccurate and we often observe large differences with respect to the reference. This feature is also observed within other closely related schemes.\cite{Maggio2017a, Wilhelm2018} As expected,\cite{Govoni2018} this issue is mostly avoided when a gKS reference is used. Certainly, using a more sophisticated algorithm to generate larger imaginary frequency grids than the present ones which are limited to a maximum of 19 points will also improve our algorithm for systems with a small KS HOMO-LUMO gap and/or low-lying core states for which generally higher resolution on the frequency axes is required.\cite{Lange2018}

Second, the PADF-approach becomes numerical unstable for large basis sets. To restore numerical stability, parts of the unoccupied space needs to be projected out during the SCF which effectively diminishes the size of the basis, especially when the basis set comprises many diffuse functions. This shortcoming can be traced back to the intrinsic difficulty to represent highly delocalized AO-pair densities using ABFs centered on two atoms only. With our auxiliary fit sets having been optimized for gKS calculations, this can lead to very large fitting coefficients which in turn cause numerical instabilities. This issue is of technical nature and can possibly be resolved by adding more diffuse functions with high angular momenta to our current auxiliary basis sets.\cite{Ihrig2015} Employing auxiliary basis sets optimized for correlated methods, as it is common practice in global DF,\cite{Weigend1998, Hattig2000, Weigend2002, Werner2003, Schutz2003, Hattig2003, Klopper2006} seems to be a promising route to approach the accuracy of canonical $G_0W_0$ also for large QZ basis sets and large systems.

Using smaller basis sets of augmented DZ and TZ quality, we calculated the HOMO and LUMO energies of a set of 250 organic molecules between 12 and 99 atoms from the GW5000 database and observed that the deviation to the FHI-AIMS CBSLE reference, not only within our scheme but also within the canonical scheme using GTO-type basis sets, is actually decreasing with increasing system size. Thus, we conclude that PADF-$G_0W_0$ calculations on the augmented DZ and TZ level can safely be performed for large systems as well. For another subset of GW5000 comprising 20 large molecules with in between 85 and 99 atoms, the aug-DZP HOMO-LUMO gap deviates by only 0.08 eV on average from the CBS limit, which is comparable to the FHI-AIMS/def2-TZVP reference.  

To summarize, it is clear that further technical improvements of our algorithm are needed. Nevertheless, the examples in this work demonstrate that already in its current form it enables accurate $G_0W_0$ calculations for large systems of hundreds of atoms with TZ and augmented DZ basis sets in a routine fashion. Not only its scalability, but also its very small prefactor make it amenable to quasi-particle and fully self-consistent GW calculations which are possible with straightforward extensions of our algorithm since we construct the complete $\Sigma$ instead of only its diagonal in the MO basis. Due to the usually consistent overestimation of QP energies, BSEs often compensate each other to a large extend in calculations of HOMO-LUMO gaps and in the past many GW calculations with augmented DZ basis sets have provided important insights into the electronic properties of practically relevant systems.\cite{Blase2011, Boulanger2014, Faber2015, Jacquemin2015, Bruneval2015, Wilhelm2016a} This indicates, that our algorithm might prove useful in practice already in its current form e.g. in the study of large organic chromophores in solution or donor-acceptor systems, and we think that its computational efficiency out-weights its current limitations to reach the CBS limit with guaranteed accuracy. 


\begin{acknowledgement}
This research received funding from the Netherlands Organisation for Scientific Research (NWO) in the framework of the Innovation Fund for Chemistry and from the Ministry of Economic Affairs in the framework of the \enquote{\emph{TKI/PPS-Toeslagregeling}}. 
\end{acknowledgement}


\begin{suppinfo}
\begin{itemize}
    \item [i)] All calculated QP HOMO and LUMO energies for the GW100 and the subsets of the GW5000 database and fit-parameters for figure~\ref{fig::s250} and additional plots.
    \item [ii)] Explicit values for thresholds controlling distance effects as well as some explanations.
    \item [iii)] .txt-file with imaginary frequency grids for some orbital energy ranges serving as starting values for our Levenberg-Marquardt algorithm as well as corresponding errors of the approximate quadratures of the MP2 energy denominator.
\end{itemize}
\end{suppinfo}


\bibliography{all}

\begin{mcitethebibliography}{249}
\bibitem[Not()]{Note-1}
The notion of $N^{x}$ always implies $\mathcal{O}\left(N^x\right)$: That is,
  the function $f$, mapping the system size to the run-time of an algorithm,
  does asymptotically not grow faster than $N^{x}$.\relax
\mciteBstWouldAddEndPunctfalse
\mciteSetBstMidEndSepPunct{\mcitedefaultmidpunct}
{}{\mcitedefaultseppunct}\relax
\EndOfBibitem
\bibitem[Not()]{Note-4}
The individual numbers of points can differ. This is due to the fact that we
  adjust the grid sizes at runtime to match a certain error parameter. Thus,
  for many systems, the number of points will actually be smaller than 18,
  since the imaginary time and frequency integrals are already converged with a
  smaller number of points.\relax
\mciteBstWouldAddEndPunctfalse
\mciteSetBstMidEndSepPunct{\mcitedefaultmidpunct}
{}{\mcitedefaultseppunct}\relax
\EndOfBibitem
\end{mcitethebibliography}


\begin{tocentry}
\includegraphics[width=\textwidth]{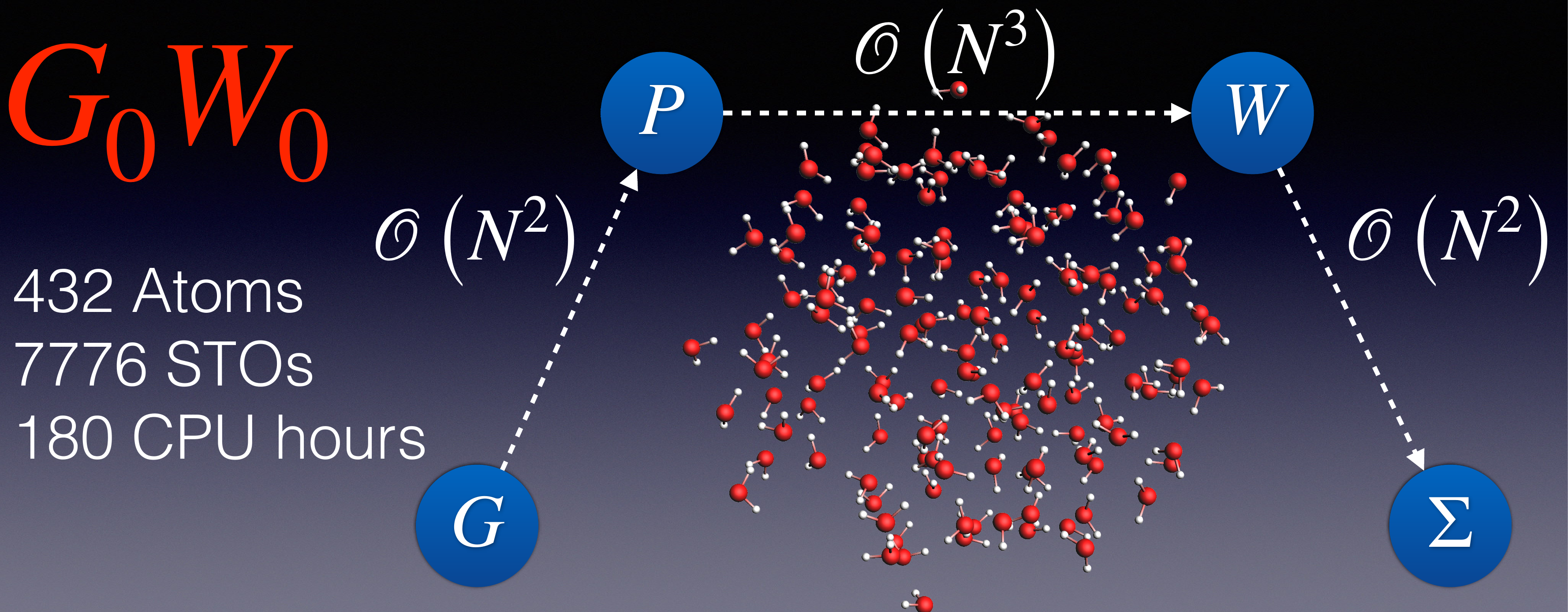}
\end{tocentry}

\includepdf[pages=-]{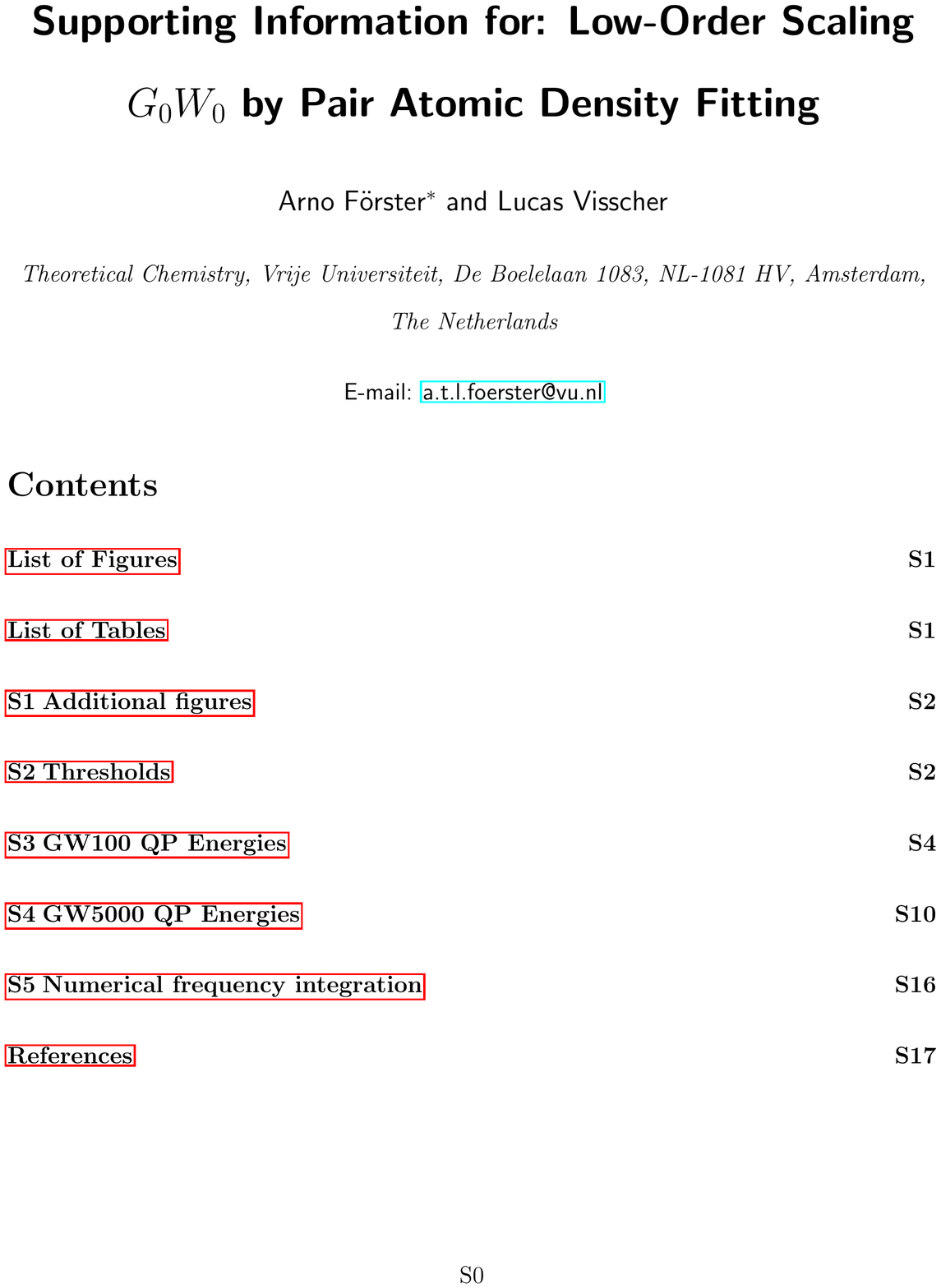}


\end{document}